\documentclass[longauth]{aa}

\usepackage[colorinlistoftodos]{todonotes}

\usepackage{graphicx}
\usepackage{txfonts}
\usepackage{hyperref}

\newcommand{\MSun}{M$_{\mathrm{\odot}}$}
\newcommand{\eg}{e.g. }
\newcommand{\ie}{i.e. }
\newcommand{\Gaia}{\textit{Gaia}}

\begin{document} 

\title{Full orbital solution for the binary system in the northern Galactic disc microlensing event Gaia16aye\footnote{Tables B.1, C.1 and D.1 are available in electronic form
at the CDS via anonymous ftp to cdsarc.u-strasbg.fr (130.79.128.5)
or via http://cdsweb.u-strasbg.fr/cgi-bin/qcat?J/A+A/}
}
\titlerunning{Full orbital solution of the Gaia16aye event}
\authorrunning{{\L}.Wyrzykowski et al.}

\author{
   {\L}ukasz Wyrzykowski\inst{\ref{inst-1}}\fnmsep\thanks{name pronunciation: {\it Woocash Vizhikovski}}
\and
P. Mr\'oz\inst{\ref{inst-1}}
\and
K. A. Rybicki\inst{\ref{inst-1}}
\and
M. Gromadzki\inst{\ref{inst-1}}
\and
Z. Ko\l{}aczkowski\inst{\ref{inst-34},\ref{inst-67}}\fnmsep\thanks{{\it deceased}}
\and
M. Zieli{\'n}ski\inst{\ref{inst-1}}
\and
P. Zieli{\'n}ski\inst{\ref{inst-1}}
\and
N. Britavskiy\inst{\ref{inst-4}, \ref{inst-4b}}
\and
A. Gomboc\inst{\ref{inst-28}}
\and
K. Sokolovsky\inst{\ref{inst-14},\ref{inst-3},\ref{inst-54}}
\and
S.T. Hodgkin\inst{\ref{inst-5}}
\and
L. Abe\inst{\ref{inst-77}}
\and
G.F. Aldi\inst{\ref{inst-15},\ref{inst-68}}
\and
A. AlMannaei\inst{\ref{inst-50},\ref{inst-91}}
\and 
G. Altavilla\inst{\ref{inst-59},\ref{inst-6}}
\and
A. Al Qasim\inst{\ref{inst-50},\ref{inst-91}}
\and
G.C. Anupama\inst{\ref{inst-7}}
\and
S. Awiphan\inst{\ref{inst-8}}
\and
E. Bachelet\inst{\ref{inst-R0}}
\and
V. Bak{\i}\c{s}\inst{\ref{inst-9}}
\and
S. Baker\inst{\ref{inst-91}}
\and
S. Bartlett\inst{\ref{inst-38}}
\and
P. Bendjoya\inst{\ref{inst-10}}
\and 
K. Benson\inst{\ref{inst-91}}
\and
I.F. Bikmaev\inst{\ref{inst-64},\ref{inst-75}}
\and
G. Birenbaum\inst{\ref{inst-11}}
\and
N. Blagorodnova\inst{\ref{inst-19}}
\and
S. Blanco-Cuaresma\inst{\ref{inst-13},\ref{inst-62}}
\and
S. Boeva\inst{\ref{inst-Bulg}}
\and
A.Z. Bonanos\inst{\ref{inst-14}}
\and
V. Bozza\inst{\ref{inst-15},\ref{inst-68}}
\and
D.M. Bramich\inst{\ref{inst-50}}
\and
I. Bruni\inst{\ref{inst-20}}
\and
R.A. Burenin\inst{\ref{inst-72},\ref{inst-73}}
\and
U. Burgaz\inst{\ref{inst-16}}
\and
T. Butterley\inst{\ref{inst-17}}
\and
H.~E. Caines\inst{\ref{inst-27}}
\and
D.~B. Caton\inst{\ref{inst-80}}
\and
S. Calchi Novati\inst{\ref{inst-71}}
\and
J.M. Carrasco\inst{\ref{inst-18}}
\and
A. Cassan\inst{\ref{inst-R2}}
\and
V. \v Cepas\inst{\ref{inst-43}}
\and
M. Cropper\inst{\ref{inst-91}}
\and
M. Chru{\'s}li\'nska\inst{\ref{inst-1},\ref{inst-19}}
\and
G. Clementini\inst{\ref{inst-20}}
\and
A. Clerici\inst{\ref{inst-28}}
\and
D. Conti\inst{\ref{inst-79b}}
\and
M. Conti\inst{\ref{inst-36}}
\and
S. Cross\inst{\ref{inst-R0}}
\and
F. Cusano\inst{\ref{inst-20}}
\and
G. Damljanovic\inst{\ref{inst-21}}
\and
A. Dapergolas\inst{\ref{inst-14}}
\and
G. D'Ago\inst{\ref{inst-69}}
\and
J.~H.~J. de Bruijne\inst{\ref{inst-22}}
\and
M. Dennefeld\inst{\ref{inst-R2}}
\and
V.~S. Dhillon\inst{\ref{inst-24},\ref{inst-4}}
\and
M. Dominik\inst{\ref{inst-25}}
\and
J. Dziedzic\inst{\ref{inst-1}}
\and
O. Erece\inst{\ref{inst-antalya}}
\and
M.~V. Eselevich\inst{\ref{inst-74}}
\and
H. Esenoglu\inst{\ref{inst-26}}
\and
L.~Eyer\inst{\ref{inst-62}}
\and
R. Figuera Jaimes\inst{\ref{inst-25},\ref{inst-R5}}
\and
S.~J. Fossey\inst{\ref{inst-27}}
\and 
A.~I. Galeev\inst{\ref{inst-64},\ref{inst-75}}
\and
S.~A. Grebenev\inst{\ref{inst-72}}
\and
A.~C. Gupta\inst{\ref{inst-90}}
\and
A.~G. Gutaev\inst{\ref{inst-64}}
\and
N. Hallakoun\inst{\ref{inst-11}}
\and
A. Hamanowicz\inst{\ref{inst-1},\ref{inst-29}}
\and 
C. Han\inst{\ref{inst-2}}
\and
B. Handzlik\inst{\ref{inst-30}}
\and
J.~B. Haislip\inst{\ref{inst-81}}
\and
L. Hanlon\inst{\ref{inst-dublin}}
\and
L.~K. Hardy\inst{\ref{inst-24}}
\and
D.~L. Harrison\inst{\ref{inst-5},\ref{inst-76}}
\and
H.J. van Heerden\inst{\ref{inst-freestate}}
\and
V.~L. Hoette\inst{\ref{inst-82}}
\and
K. Horne\inst{\ref{inst-25}}
\and
R. Hudec\inst{\ref{inst-31}, \ref{inst-64}, \ref{inst-31b}}
\and
M. Hundertmark\inst{\ref{inst-R4}}
\and
N. Ihanec\inst{\ref{inst-28}}
\and 
E.~N. Irtuganov\inst{\ref{inst-64},\ref{inst-75}}
\and
R. Itoh\inst{\ref{inst-32}}
\and
P. Iwanek\inst{\ref{inst-1}}
\and
M.D.Jovanovic\inst{\ref{inst-21}}
\and
R. Janulis\inst{\ref{inst-43}}
\and
M. Jel\'inek\inst{\ref{inst-31}}
\and
E. Jensen\inst{\ref{inst-79}}
\and
Z. Kaczmarek\inst{\ref{inst-1}}
\and
D. Katz\inst{\ref{inst-92}}
\and
I.M. Khamitov\inst{\ref{inst-33},\ref{inst-64}}
\and
Y.Kilic\inst{\ref{inst-antalya}}
\and
J. Klencki\inst{\ref{inst-1},\ref{inst-19}}
\and
U. Kolb\inst{\ref{inst-35b}}
\and
G. Kopacki\inst{\ref{inst-34}}
\and
V.~V.~Kouprianov\inst{\ref{inst-81}}
\and
K. Kruszy\'nska\inst{\ref{inst-1}}
\and
S. Kurowski\inst{\ref{inst-30}}
\and
G. Latev\inst{\ref{inst-Bulg}}
\and
C-H. Lee\inst{\ref{inst-chlee}, \ref{inst-chlee-b}}
\and
S. Leonini\inst{\ref{inst-36}}
\and
G. Leto\inst{\ref{inst-37}}
\and
F. Lewis\inst{\ref{inst-38},\ref{inst-47}}
\and
Z. Li\inst{\ref{inst-R0}}
\and
A. Liakos\inst{\ref{inst-14}}
\and
S.~P. Littlefair\inst{\ref{inst-24}}
\and
J. Lu\inst{\ref{inst-39}}
\and
C.J. Manser\inst{\ref{inst-40}}
\and
S. Mao\inst{\ref{inst-R5}}
\and
D. Maoz\inst{\ref{inst-11}}
\and
A.Martin-Carrillo\inst{\ref{inst-dublin}}
\and
J.~P. Marais\inst{\ref{inst-freestate}}
\and
M. Maskoli\={u}nas\inst{\ref{inst-43}}
\and
J.~R. Maund\inst{\ref{inst-24}}
\and 
P.~J. Meintjes\inst{\ref{inst-freestate}}
\and
S.~S. Melnikov\inst{\ref{inst-64},\ref{inst-75}}
\and
K. Ment\inst{\ref{inst-R4}}
\and
P. Miko{\l}ajczyk\inst{\ref{inst-34}}
\and
M. Morrell\inst{\ref{inst-35b}}
\and
N. Mowlavi\inst{\ref{inst-62}}
\and
D. Mo{\'z}dzierski\inst{\ref{inst-34}}
\and
D. Murphy\inst{\ref{inst-dublin}}
\and
S. Nazarov\inst{\ref{inst-78}}
\and
H. Netzel\inst{\ref{inst-1},\ref{inst-67}}
\and
R. Nesci\inst{\ref{inst-55}}
\and
C.-C. Ngeow\inst{\ref{inst-41}}
\and
A.~J. Norton\inst{\ref{inst-35b}}
\and
E. O. Ofek\inst{\ref{inst-42}}
\and
E. Pak\v stien\.e\inst{\ref{inst-43}}
\and
L. Palaversa\inst{\ref{inst-5},\ref{inst-62}}
\and
A. Pandey\inst{\ref{inst-90}}
\and
E. Paraskeva\inst{\ref{inst-14},\ref{inst-66}}
\and
M. Pawlak\inst{\ref{inst-1},\ref{inst-53}}
\and
M.~T. Penny\inst{\ref{inst-45}}
\and
B.~E. Penprase\inst{\ref{inst-46}}
\and
A. Piascik\inst{\ref{inst-47}}
\and
J.~L.~Prieto\inst{\ref{inst-83}, \ref{inst-83b}}
\and
J.~K.~T. Qvam\inst{\ref{inst-84}}
\and
C. Ranc\inst{\ref{inst-R7}}
\and
A. Rebassa-Mansergas\inst{\ref{inst-48},\ref{inst-58}}
\and
D.~E.~Reichart\inst{\ref{inst-81}}
\and
P. Reig\inst{\ref{inst-49},\ref{inst-63}}
\and
L. Rhodes\inst{\ref{inst-24}}
\and
J.-P. Rivet\inst{\ref{inst-77}}
\and
G. Rixon\inst{\ref{inst-5}}
\and
D. Roberts\inst{\ref{inst-35b}}
\and
P. Rosi\inst{\ref{inst-36}}
\and
D.M. Russell\inst{\ref{inst-50}}
\and
R. Zanmar Sanchez\inst{\ref{inst-37}}
\and
G. Scarpetta\inst{\ref{inst-15},\ref{inst-70}}
\and
G. Seabroke\inst{\ref{inst-91}}
\and
B.~J. Shappee\inst{\ref{inst-57}}
\and
R. Schmidt\inst{\ref{inst-R4}}
\and
Y. Shvartzvald\inst{\ref{inst-11b},\ref{inst-11c}}
\and
M. Sitek\inst{\ref{inst-1}}
\and
J. Skowron\inst{\ref{inst-1}}
\and
M. \'Sniegowska\inst{\ref{inst-1}, \ref{inst-65}, \ref{inst-67}}
\and
C. Snodgrass\inst{\ref{inst-35}}
\and
P.~S. Soares\inst{\ref{inst-27}}
\and
B. van Soelen\inst{\ref{inst-freestate}}
\and
Z.~T. Spetsieri\inst{\ref{inst-14},\ref{inst-66}}
\and
A. Stankevi\v{c}i\={u}t\.{e}\inst{\ref{inst-1}}
\and
I.~A. Steele\inst{\ref{inst-47}}
\and
R.~A. Street\inst{\ref{inst-R0}}
\and
J. Strobl\inst{\ref{inst-31}}
\and
E. Strubble\inst{\ref{inst-82}}
\and
H. Szegedi\inst{\ref{inst-freestate}}
\and
L.~M. Tinjaca Ramirez\inst{\ref{inst-36}}
\and
L. Tomasella\inst{\ref{inst-52}}
\and
Y. Tsapras\inst{\ref{inst-R4}}
\and
D. Vernet\inst{\ref{inst-10}}
\and
S. Villanueva Jr.\inst{\ref{inst-45}}
\and
O. Vince\inst{\ref{inst-21}}
\and
J. Wambsganss\inst{\ref{inst-R4}, \ref{inst-R4b}}
\and
I. P. van der Westhuizen\inst{\ref{inst-freestate}}
\and
K. Wiersema\inst{\ref{inst-40},\ref{inst-56}}
\and
D. Wium\inst{\ref{inst-freestate}}
\and
R.~W. Wilson\inst{\ref{inst-17}}
\and 
A. Yoldas\inst{\ref{inst-5}}
\and
R.Ya. Zhuchkov\inst{\ref{inst-64},\ref{inst-75}}
\and
D.~G. Zhukov\inst{\ref{inst-64}} 
\and 
J. Zdanavi\v cius\inst{\ref{inst-43}}
\and
S. Zo\l{}a\inst{\ref{inst-30},\ref{inst-30b}}
\and
A. Zubareva\inst{\ref{inst-60},\ref{inst-3}}
}
\institute{
 Warsaw University Astronomical Observatory, Al. Ujazdowskie 4, 00-478 Warszawa, Poland\label{inst-1}
\and
Department of Physics, Chungbuk National University, Cheongju 28644, Republic of Korea
\label{inst-2}
\and
Sternberg Astronomical Institute, Moscow State University, Universitetskii~pr.~13, 119992~Moscow, Russia\label{inst-3}
\and
 Instituto de Astrofisica de Canarias (IAC), E-38205 La Laguna, Tenerife, Spain\label{inst-4}
\and
Universidad de La Laguna, Dpto. Astrof{\'i}sica, E-38206 La Laguna, Tenerife, Spain\label{inst-4b}
\and
 Institute of Astronomy, University of Cambridge, Madingley Road CB3 0HA, Cambridge, UK\label{inst-5}
\and
 INAF - Osservatorio Astronomico di Roma, Via di Frascati 33, 00078 Monte Porzio Catone (Roma), Italy\label{inst-6}
\and
 Indian Institute of Astrophysics, II Block Koramangala, Bengaluru 560034, India\label{inst-7}
\and
National Astronomical Research Institute of Thailand, 260, Moo 4, T. Donkaew, A. Mae Rim, Chiang Mai, 50180, Thailand \label{inst-8}
\and
Department of Space Sciences and Technologies, Faculty of Science, Akdeniz University, 07058, Antalya, Turkiye\label{inst-9}
\and
Universit\'e C\^{o}te d'Azur, Observatoire de la C\^{o}te d'Azur, CNRS, Laboratoire Lagrange, France\label{inst-10}
\and 
School of Physics and Astronomy, Tel-Aviv University, Tel-Aviv 6997801, Israel\label{inst-11}
\and
Jet Propulsion Laboratory, California Institute of Technology, 4800 Oak Grove Drive, Pasadena, CA 91109, USA\label{inst-11b}
\and
NASA Postdoctoral Program Fellow\label{inst-11c}
\and
Harvard-Smithsonian Center for Astrophysics, 60 Garden Street, Cambridge, MA 02138, USA\label{inst-13}
\and
  Institute of Astronomy and NAO Rozhen, BAS, 72 Tsarighradsko Shousse Blvd., 1784 Sofia, Bulgaria\label{inst-Bulg}
  \and
  National Optical Astronomy Observatory 950 N Cherry Avenue, Tucson, AZ 85719, USA\label{inst-chlee}
  \and
  Subaru Telescope, National Astronomical Observatory of Japan, 650 N Aohoku Place, Hilo, HI 96720, USA\label{inst-chlee-b}
  \and
IAASARS, National Observatory of Athens, Vas.~Pavlou \& I.~Metaxa, 15236~Penteli, Greece \label{inst-14}
\and
Dipartimento di Fisica E.R. Caianiello, Universit\`a di Salerno, Via Giovanni Paolo II 132, I-84084 Fisciano (SA), Italy\label{inst-15}
\and
Department of Astronomy and Space Sciences, Ege University, 35100 Izmir, Turkey\label{inst-16}
\and
 Centre for Advanced Instrumentation, University of Durham, South Road, Durham DH1 3LE, United Kingdom\label{inst-17}
\and
 Institut del Ci\`encies del Cosmos (ICC), Universitat de Barcelona (IEEC-UB), c/ Mart\'{\i} i Franqu\`es, 1, 08028 Barcelona, Spain \label{inst-18}
\and
 Department of Astrophysics/IMAPP, Radboud University Nijmegen, P.O. Box 9010, 6500 GL Nijmegen, The Netherlands \label{inst-19}
\and
 INAF - Osservatorio di Astrofisica e Scienza dello Spazio di Bologna, via Gobetti 93/3 - 40129 Bologna - Italy\label{inst-20}
\and
Astronomical Observatory, Volgina 7, 11060 Belgrade, Serbia \label{inst-21}
\and
 Science Support Office, Directorate of Science, European Space Research and Technology Centre (ESA/ESTEC), Keplerlaan 1, 2201 AZ, Noordwijk, The Netherlands\label{inst-22}
\and
Qatar Environment and Energy Research Institute(QEERI), HBKU, Qatar Foundation, Doha, Qatar\label{inst-R1}
\and
Institut d'Astrophysique de Paris, Sorbonne Universit{\'e}, CNRS, UMR 7095, 98 bis bd Arago, 75014 Paris, France\label{inst-R2}
\and
Department of Physics and Astronomy, University of Sheffield, Sheffield S3 7RH, UK\label{inst-24}
\and
 Centre for Exoplanet Science, SUPA School of Physics \& Astronomy, University of St Andrews, North Haugh, St Andrews, KY16 9SS, United Kingdom\label{inst-25}
\and
Akdeniz University, Dumlupinar Blv., Campus, 07058, Antalya, Turkey\label{inst-antalya}
\and
Istanbul University, Department of Astronomy and Space Sciences, 34119 Beyazit, Istanbul, Turkey\label{inst-26}
\and
 Dept. of Physics \& Astronomy, UCL, Gower St., London WC1E 6BT, UK\label{inst-27}
\and
 Center for Astrophysics and Cosmology, University of Nova Gorica, Vipavska cesta 11c, 5270 Ajdov\v s\v cina, Slovenia\label{inst-28}
\and
 European Southern Observatory, Karl Schwarzschild Str 2, D-85748 Garching, Germany\label{inst-29}
\and
Astronomical Observatory, Jagiellonian University, Krak{\'o}w, Poland\label{inst-30}
\and
Mt. Suhora Observatory, Pedagogical University, ul. Podchor{\,a}{\.z}ych 2, 30-084 Krak{\'o}w, Poland\label{inst-30b}
\and
 Astronomical Institute of the Academy of Sciences of the Czech Republic, Ond\v rejov, Czech Republic\label{inst-31}
\and
Czech Technical University, Faculty of Electrical Engineering, Technick{\'a} 2. 166 27 Praha 6 , Czech Republic\label{inst-31b}
\and
Zentrum f{\"u}r Astronomie der Universit{\"a}t Heidelberg, Astronomisches Rechen-Institut, M{\"o}nchhofstr. 12-14, 69120 Heidelberg, Germany\label{inst-R4}
\and
International Space Science Institute, Hallerstrasse 6, CH-3012 Bern, Switzerland\label{inst-R4b}
\and
 Department of Physics, School of Science, Tokyo Institute of Technology, 2-12-1 Ohokayama, Meguro, Tokyo 152-8551, Japan\label{inst-32}
\and
 T\"UB\.ITAK National Observatory, Akdeniz University Campus, 07058 Antalya, Turkey\label{inst-33}
\and
Instytut Astronomiczny Uniwersytetu Wroc{\l}awskiego, ul. Kopernika 11, 51-622 Wroc{\l}aw, Poland\label{inst-34}
\and
Institute for Astronomy, University of Edinburgh, Royal Observatory, Edinburgh EH9 3HJ, UK\label{inst-35}
\and
School of Physical Sciences, The Open University, Walton Hall, Milton Keynes MK7 6AA, UK\label{inst-35b}
\and
Osservatorio Astronomico Provinciale di Montarrenti, S. S. 73 Ponente, I-53018, Sovicille, Siena, Italy\label{inst-36}
\and
 INAF - Osservatorio Astrofisico di Catania, Via Santa Sofia 78, I-95123 Catania, Italy\label{inst-37}
\and
Faulkes Telescope Project, School of Physics, and Astronomy, Cardiff University, The Parade, Cardiff CF24 3AA, UK\label{inst-38}
\and
 Astronomy Department, University of California, Berkeley, CA 94720, USA\label{inst-39}
\and
 Department of Physics, University of Warwick, Coventry CV4 7AL, UK\label{inst-40}
\and
National Astronomical Observatories, Chinese Academy of Sciences, 100012 Beijing, China\label{inst-R5}
\and
 Graduate Institute of Astronomy, National Central University, Jhongli 32001, Taiwan\label{inst-41}
\and
Department of particle physics and astrophysics, Weizmann Institute of Science, Revovot, Israel\label{inst-42}
\and
 Institute of Theoretical Physics and Astronomy, Vilnius University, Saul\.etekio av. 3, 10257 Vilnius, Lithuania\label{inst-43}
\and
 Department of Astronomy, Ohio State University, 140 W. 18th Ave., Columbus, OH 43210, USA\label{inst-45}
\and
 Soka University of America, 1 University Drive, Aliso Viejo, CA 92656, USA\label{inst-46}
\and
 Astrophysics Research Institute, Liverpool John Moores University, 146 Brownlow Hill, Liverpool L3 5RF, UK\label{inst-47}
\and
 Universitat Polit\`ecnica de Catalunya, Departament de F\'\i sica, c/Esteve Terrades 5, 08860 Castelldefels, Spain\label{inst-48}
\and
 Institute of Astrophysics, Foundation for Research and Technology-Hellas, 71110 Heraklion, Crete, Greece\label{inst-49}
\and
 New York University Abu Dhabi, Saadiyat Island, Abu Dhabi, P.O. Box 129188, United Arab Emirates\label{inst-50}
\and
 Las Cumbres Observatory Global Telescope Network, 6740 Cortona Drive, suite 102, Goleta, CA 93117, USA\label{inst-R0}
\and
INAF Osservatorio Astronomico di Padova, Vicolo dell'Osservatorio 5, I-35122 Padova, Italy\label{inst-52}
\and
 Institute of Theoretical Physics, Faculty of Mathematics and Physics, Charles University in Prague, Czech Republic\label{inst-53}
\and
Astro Space Center of Lebedev Physical Institute, Profsoyuznaya~St.~84/32, 117997~Moscow, Russia\label{inst-54}
\and
INAF - Istituto di Astrofisica e Planetologia Spaziali, Roma, Italy\label{inst-55}
 \and 
Department of Physics and Astronomy, University of Leicester, University Road, Leicester LE1 7RH, UK \label{inst-56}
\and
Institute for Astronomy, University of Hawai'i, 2680 Woodlawn Drive, Honolulu, HI 96822, USA \label{inst-57}
\and
Astrophysics Science Division, NASA/Goddard Space Flight Center, Greenbelt, MD 20771, USA\label{inst-R7}
\and
Institute for Space Studies of Catalonia, c/Gran Capit{\'a} 2--4, Edif. Nexus 104, 08034 Barcelona, Spain \label{inst-58}
\and
Space Science Data Center - ASI, Via del Politecnico SNC, 00133 Roma, Italy  \label{inst-59}
\and
Institute of Astronomy, Russian Academy of Sciences, Pyatnitskaya str. 48, 119017 Moscow, Russia \label{inst-60}
\and
Observatoire de Gen\`eve, Universit\'e de Gen\`eve, CH-1290 Versoix, Switzerland\label{inst-62}
\and 
University of Crete, Physics Department \& Institute of Theoretical \& Computational Physics, 71003 Heraklion, Crete, Greece\label{inst-63}
\and 
Kazan Federal University, ul. Kremlevskaya 18, 420008 Kazan, Russia\label{inst-64}
\and 
Center for Theoretical Physics,Polish Academy of Sciences, Al. Lotnik\' ow 32/46,02-668 Warsaw, Poland\label{inst-65}
\and 
Department of Astrophysics, Astronomy \& Mechanics, Faculty of Physics, University of Athens, 15783 Athens, Greece\label{inst-66}
\and 
Nicolaus Copernicus Astronomical Center, Polish Academy of Sciences, ul. Bartycka 18, 00-716 Warsaw, Poland\label{inst-67}
\and 
Istituto Nazionale di Fisica Nucleare, Sezione di Napoli, Italy\label{inst-68}
\and 
Instituto de Astrof{\'i}sica, Facultad de F{\'i}sica, Pontificia Universidad Cat{\'o}lica de Chile, Av. Vicu{\~n}a Mackenna 4860, 7820436 Macul, Santiago, Chile
\label{inst-69}
\and 
Istituto Internazionale per gli Alti Studi Scientifici (IIASS), Via G. Pellegrino 19, I-84019 Vietri sul Mare (SA), Italy\label{inst-70}
\and 
IPAC, Mail Code 100-22, California Institute of Technology, 1200 East California Boulevard, Pasadena, CA 91125, USA\label{inst-71}
\and 
Space Research Institute of Russian Academy of Sciences (IKI),  84/32 Profsoyuznaya, Moscow, Russia\label{inst-72}
\and 
National Research University Higher School of Economics, Myasnitskaya ul. 20, 101000 Moscow, Russia\label{inst-73}
\and 
Institute of Solar-Terrestrial Physics SB RAS, Irkutsk, Russia\label{inst-74}
\and
Academy of Sciences of Tatarstan, Kazan, Russia\label{inst-75}
\and
Kavli Institute for Cosmology, Madingley Road, Cambridge, CB3 0HA, United Kingdom\label{inst-76}
\and
Universit\'e C\^ote d'Azur, OCA, CNRS, Laboratoire Lagrange, Nice, France\label{inst-77}
\and
Crimean Astrophysical Observatory, Nauchnyi, Crimea\label{inst-78}
\and
It should be: American Association of Variable Star Observers (AAVSO), 49 Bay State Road, Cambridge, MA 02138, USA\label{inst-79b}
\and
Swarthmore College, 500 College Avenue, Swarthmore, PA 19081, USA\label{inst-79}
\and
Dark Sky Observatory, Department of Physics and Astronomy, Appalachian State University, Boone, NC 28608, USA\label{inst-80}
\and
University of North Carolina at Chapel Hill, Chapel Hill, North Carolina NC 27599, USA\label{inst-81}
\and
Yerkes Observatory, Department of Astronomy and Astrophysics, University of Chicago, 373 W. Geneva St., Williams Bay, WI 53191, USA\label{inst-82}
\and
N{\'u}cleo de Astronom{\'i}a de la Facultad de Ingenier{\'i}a, Universidad Diego Portales, Av. Ej{\'e}rcito 441, Santiago, Chile
\label{inst-83}
\and
Millenium Institute of Astrophysics, Santiago, Chile 
\label{inst-83b}
\and
Horten Upper Secondary School, Bekkegata 2, 3181 Horten, Norway\label{inst-84}
\and
  Aryabhatta Research Institute of Observational Sciences (ARIES), Manora Peak, Nainital - 263002, India\label{inst-90}
\and
Mullard Space Science Laboratory, University College London, Holmbury St Mary, Dorking, RH5 6NT, UK
\label{inst-91}
\and
GEPI, Observatoire de Paris, Universit{\'e} PSL, CNRS, 5 Place Jules Janssen, 92190 Meudon, France
\label{inst-92}
\and
School of Physics, University College Dublin, Belfield, Dublin 4, Ireland
\label{inst-dublin}
\and
Department of Physics, Faculty of Natural and Agricultural Sciences, University of the Free State, Bloemfontein 9300, Republic of South Africa
\label{inst-freestate}
}
   \date{Received }

  \abstract{
   Gaia16aye was a binary microlensing event discovered in the direction towards the northern Galactic disc and was one of the first microlensing events detected and alerted to by the {\Gaia} space mission. Its light curve exhibited five distinct brightening episodes, reaching up to I=12 mag, and it was covered in great detail with almost 25,000 data points gathered by a network of telescopes. 
   We present the photometric and spectroscopic follow-up covering 500 days of the event evolution. We employed a full Keplerian binary orbit microlensing model combined with the motion of Earth and {\Gaia} around the Sun to reproduce the complex light curve. 
   The photometric data allowed us to solve the microlensing event entirely and to derive the complete and unique set of orbital parameters of the binary lensing system. 
   We also report on the detection of the first-ever microlensing space-parallax between the Earth and {\Gaia} located at L2.  
The properties of the binary system were derived from microlensing parameters, and we found that the system is composed of two main-sequence stars with masses 0.57$\pm$0.05 \MSun \, and 0.36$\pm$0.03 \MSun \, at 780 pc, with an orbital period of 2.88 years and an eccentricity of 0.30.
We also predict the astrometric microlensing signal for this binary lens as it will be seen by {\Gaia} as well as the radial velocity curve for the binary system.
   Events such as Gaia16aye indicate the potential for the microlensing method of probing the mass function of dark objects, including black holes, in directions other than that of the Galactic bulge.
   This case also emphasises the importance of long-term time-domain coordinated observations that can be made with a network of heterogeneous telescopes. 
} 

   \maketitle
%

\section{Introduction}
Measuring the masses of stars or stellar remnants is one of the most challenging tasks in modern astronomy. 
Binary systems were the first to facilitate mass measurement through the Doppler effect in radial velocity measurements
\citep[\eg][]{Popper1967}, leading to the mass-luminosity relation and an advancement in the understanding of stellar evolution \citep[\eg][]{Paczynski1971, 2010Pietrzynski}. 
However, these techniques require the binary components to emit detectable amounts of light, often demanding large-aperture telescopes and sensitive instruments. 
In order to study the invisible objects, in particular stellar remnants such as neutron stars or black holes, other means of mass measurement are necessary. 
Recently, the masses of black holes were measured when a close binary system tightened its orbit and emitted gravitational waves \citep[\eg][]{Abbott2016}, yielding unexpectedly high masses that were not observed before \citep[\eg][]{Abbott2017-50msunBH, Belczynski2016, Bird2016}.
Because of the low merger rates, gravitational wave experiment detections are limited to very distant galaxies. Other means of mass measurement are therefore required to probe the faint and invisible populations in the Milky Way and its vicinity.

Gravitational microlensing allows for detection and study of binary systems regardless of the amount of light they emit and regardless of the radial velocities of the components, 
as long as the binary crosses the line of sight to a star that is bright enough to be observed. Therefore, this method offers an opportunity to detect binary systems that contain 
planets \citep[\eg][]{GouldLoeb1992, Albrow1998, Bond2004, Udalski2005}, planets orbiting a binary system of stars \citep[\eg][]{Poleski2014, Bennett2016}, and 
black holes or other dark stellar remnants \citep[\eg][]{Shvartzvald2015}.

Typically, searches for microlensing events are conducted in the direction of the Galactic bulge because of the high stellar density, potential sources and lenses, and the high microlensing optical depth \citep[\eg][]{KiragaPaczynski1994, Udalski1994, Paczynski1996, Wozniak2001, Sumi2013, Udalski2015, Wyrzykowski2015, Mroz2017}. The regions of the Galactic plane outside of the bulge have occasionally also been monitored in the past for microlensing events, however, even though the predicted rates of events were orders of magnitude lower \citep[\eg][]{Han2008, Gaudi2008}. 
\cite{Derue2001} first published microlensing events that were detected during the long-term monitoring of the selected disc fields. Two serendipitous discoveries of bright microlensing events outside of the bulge were reported by amateur observers, the Tago event \citep{Fukui2007, Gaudi2008}, and the Kojima-1 event \citep{Nucita2018, Dong2019, Fukui2019}, which has a signature of a planet next to the lens. 
The first binary microlensing event in the Galactic disc was reported in \cite{Rahal2009} (GSA14), but its light curve was too poorly sampled in order to conclude on the parameters of the binary lens. 

The best-sampled light curves come from bulge surveys, such as MACHO \citep{Alcock1997, Popowski2001}, the Exp\'erience pour la Recherche d'Objets Sombres (EROS) \citep{Hamadache2006}, the Optical Gravitational Lensing Experiment (OGLE) \citep{Udalski1994, Udalski2000, Udalski2015}, the Microlensing Observations in Astrophysics (MOA) \citep{Yock1998, Sumi2013}, and  the Korean Microlensing Telescope Network (KMNet) \citep{KMTNET}. In particular, the OGLE project has been monitoring the Galactic bulge regularly since 1992 and was the first to report on a binary microlensing event in 1993 \citep{Udalski1994bin}. 
Binary microlensing events constitute about 10\% of all events reported by the microlensing surveys of the bulge.
The binary lens differs from a single lens when the component separation on the sky is of order of their Einstein radius \citet{Paczynski1996, Gould2000b}, which is computed as
\begin{equation}
\label{eq:thetaE}
\theta_E=\sqrt{\kappa M_L (\pi_{\rm l} - \pi_{\rm s})}, ~~~\kappa \equiv \frac{4G}{c^2} \approx 8.144~\mathrm{mas~M_\odot^{-1}},\end{equation}
where $M_L$ is the total mass of the binary and $\pi_{\rm l}$ and $\pi_{\rm s}$ are parallaxes of the lens and the source, respectively. 
For the conditions in the Galaxy and a typical mass of the lens, the size of the Einstein ring is about 1 milliarcsecond (1\,mas).
%
Instead of a circular Einstein ring as in the case of a single lens (or very tight binary system), two (or more) lensing objects produce a complex curve on the sky, shaped by the mass ratio and projected separation of the components. This is called the critical curve.
In the source plane this curve turns into a caustic curve (as opposed to a point in the case of a single lens), which denotes the places where the source is infinitely amplificated  \citep[\eg][]{Bozza2001, Rattenbury2009}.
As the source and the binary lens move, their relative proper motion changes the position of the source with respect to the caustics. Depending on this position, there are three
(when the source is outside of the caustic) or five (inside the caustic) images of the source. 
Images also change their location as well as their size, therefore the combined light of the images we observe changes the observed amplification, with the most dramatic changes at the caustic crossings. 
In a typical binary lensing event the source--lens trajectory can be approximated with a straight line  \citep[\eg][]{Jaroszynski2004, Skowron2007}. If the line crosses the caustic, it produces a characteristic U-shaped light curve because the amplification increases steeply as the source approaches the caustic and remains high inside the caustic  \citep[\eg][]{Witt1995}. If the source approaches one of the caustic cusps, the light curve shows a smooth increase, similar to a single lensing event. Identifying all these features in the light curve helps constrain the shape of the caustic and hence the parameters of the binary. 
An additional annual parallax effect causes the trajectory of the source to curve,
which probes the caustic shape at multiple locations  \citep[\eg][]{An2001, SkowronWyrzykowski2009, Udalski2018Quintuple}
and thus helps constrain the solution of the binary system better.

The situation becomes more complex when a binary system rotates while lensing, which causes the binary configuration on the sky to change. This in turn changes the shape and size of the caustic \citep{Albrow2000}. 
In the case of most binary microlensing events the effect of the orbital motion can be neglected because the orbital periods are often much longer (typically years) than the duration of the event (typically weeks). 
However, in longer events the orbital motion has to be taken into account in the model. 
Together with the source--lens relative motion and the parallax effect, this causes the observed amplification to vary significantly during the event and may generate multiple crossings of the caustic and amplification due to cusp approach \citep[\eg][]{SkowronWyrzykowski2009}. 
However, in rare cases, such a complex event allows us not only to measure the mass and distance of the lens, but also to derive all orbital parameters of the binary. 
The first such case was found by the OGLE survey in the event OGLE-2009-BLG-020 \citep{Skowron2011}, and its orbital parameters found in the model were verified with radial velocity measurement  \citep{Yee2016RV}. 
The orbital motion was also modelled in the MOA-2011-BLG-090 and OGLE-2011-BLG-0417 events \citep{shin2012}, but the former was too faint and the latter was not confirmed with radial velocity data \citep{Boisse2015RV, Bachelet2018}. 


Additional information that helps constrain the parameters of the system may also come from space parallax \citep[\eg][]{Refsdal1966, Gould1992, Gould2009}.
This is now being routinely done by observing microlensing events from the Earth and \textit{Spitzer} or \textit{Kepler}, separated by more than 1 au \citep[\eg][]{Udalski2015Spitzer, Yee2015, CalchiNovati2016, Shvartzvald2016, Zhu2016, Poleski2016}.

The most difficult parameter to measure, however, is the size of the Einstein radius.
It can be found when the finite source effects are detected, when the angular source size is large enough to experience a significant gradient in the magnification near the centre of the Einstein ring or the binary lens caustic \citep[\eg][]{Yoo2004, Zub2011}.
The measurement of the angular separation between the luminous lens and the source years or decades after the event also directly leads to calculation of $\theta_E$  \citep[\eg][]{Kozlowski2007}.
Otherwise, for dark lenses, the measurement of $\theta_E$ can only come from astrometric microlensing \citep{DominikSahu2000, Belokurov2002, Lu2016, Kains2017, Sahu2017}.
As shown in \cite{Rybicki2018}, {\Gaia} will soon provide precise astrometric observations for microlensing events,  which will allow us to measure $\theta_E$, but only for events brighter than about $V < 15$ mag.

Here we present Gaia16aye, a unique event from the Galactic disc, far from the Galactic bulge, which lasted almost two years and exhibited effects of binary lens rotation, an annual and space parallax, and a finite source. 
The very densely sampled light curve was obtained solely thanks to an early alert from {\Gaia} and a dedicated ground-based follow-up of tens of observers, including amateurs and school pupils. 
The wealth of photometric data allowed us to find the unique solution for the binary system parameters. 

The paper is organised as follows. 
Sections 2 and 3 describe the history of the detection and the photometric and spectroscopic data collected during the follow-up of Gaia16aye. 
In Section 4 we describe the microlensing model we used to reproduce the data.
We then discuss the results in Section 5.


\begin{figure*}[ht]
\begin{center}
\includegraphics[width=0.92\textwidth]{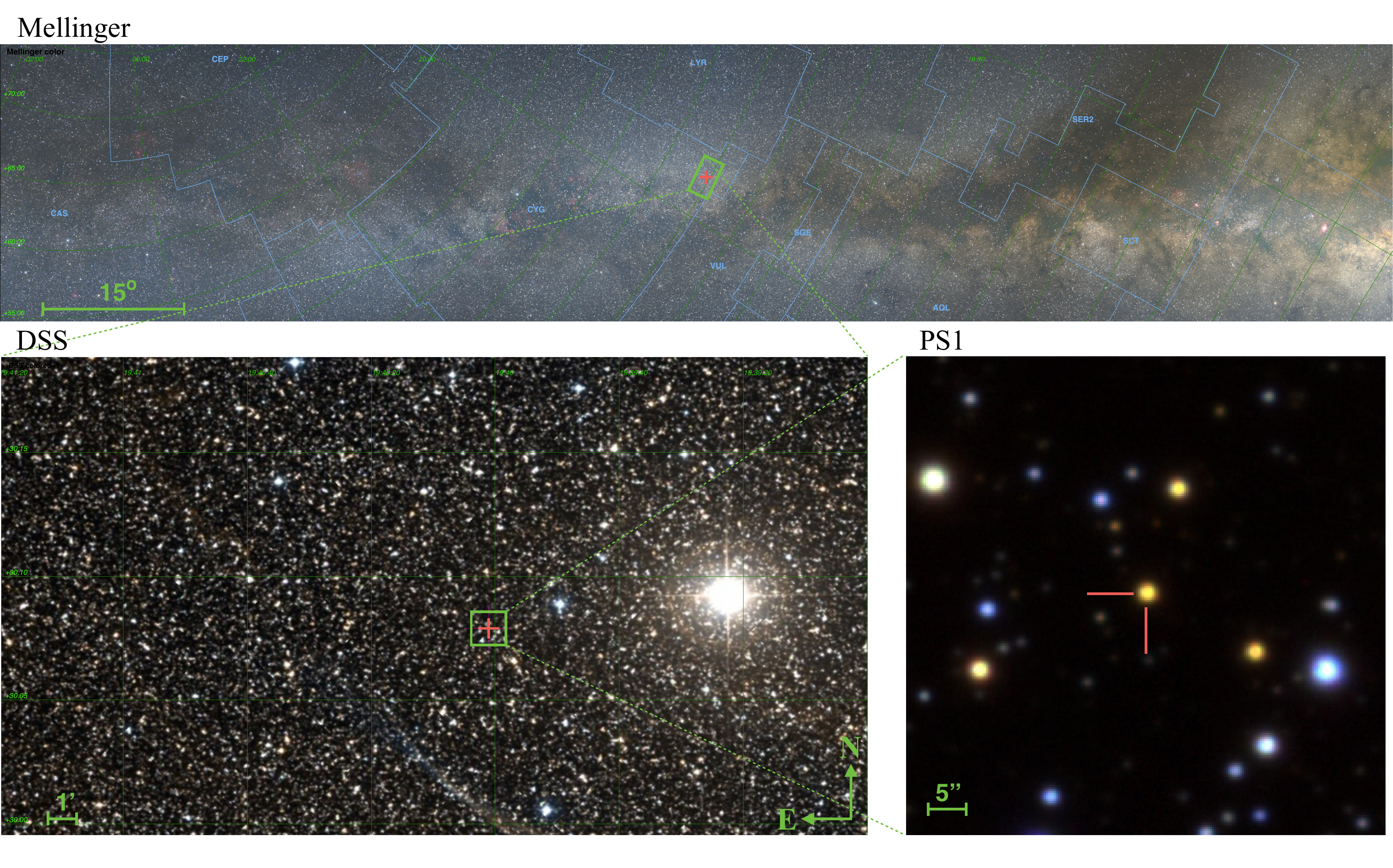}
\caption{Location of Gaia16aye on the sky. Images from Mellinger and DSS were obtained using the Aladin tool.}
\label{fig:finding}
\end{center}
\end{figure*}

\begin{figure*}[h]
\begin{center}
\includegraphics[width=\textwidth]{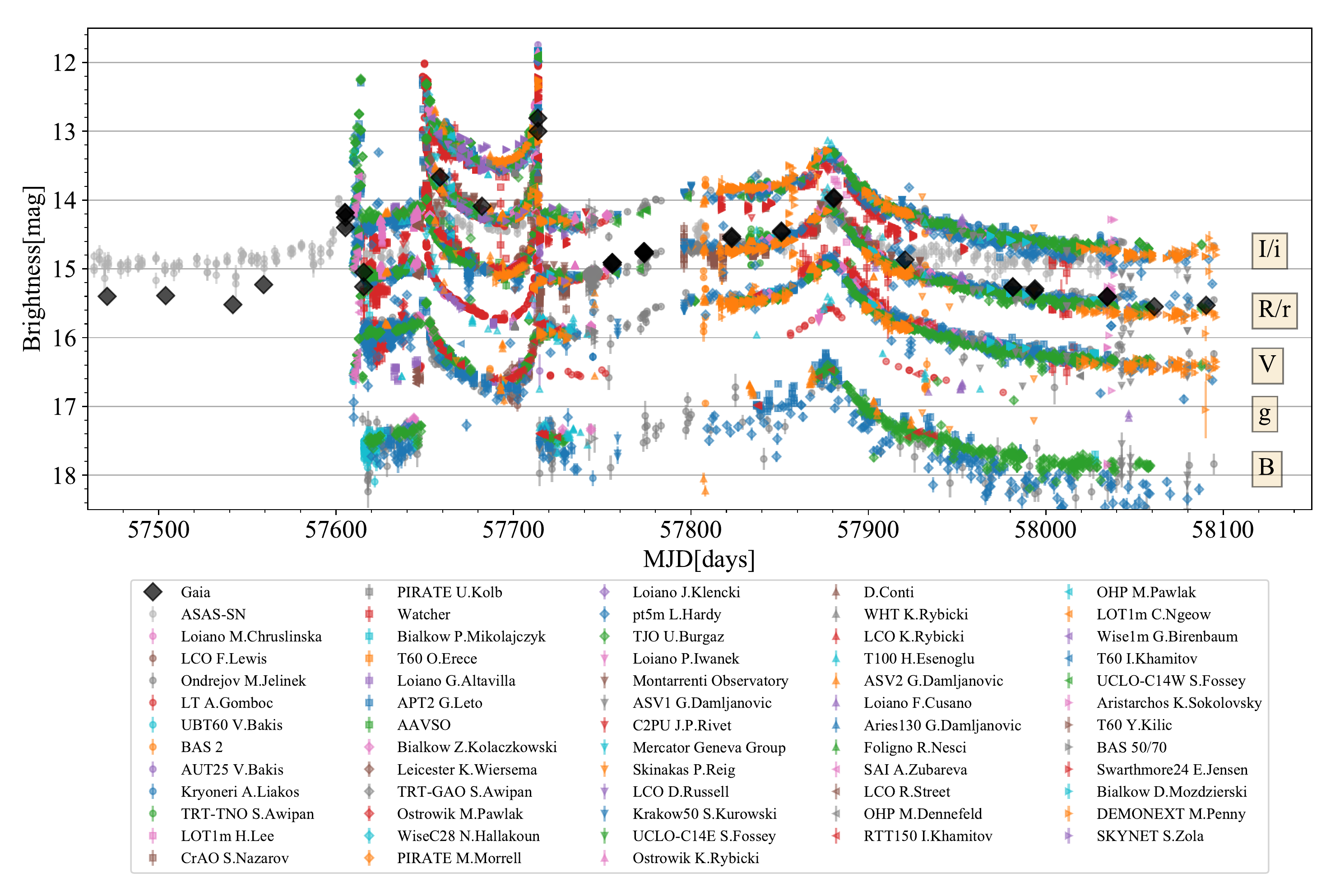}
\includegraphics[width=\textwidth]{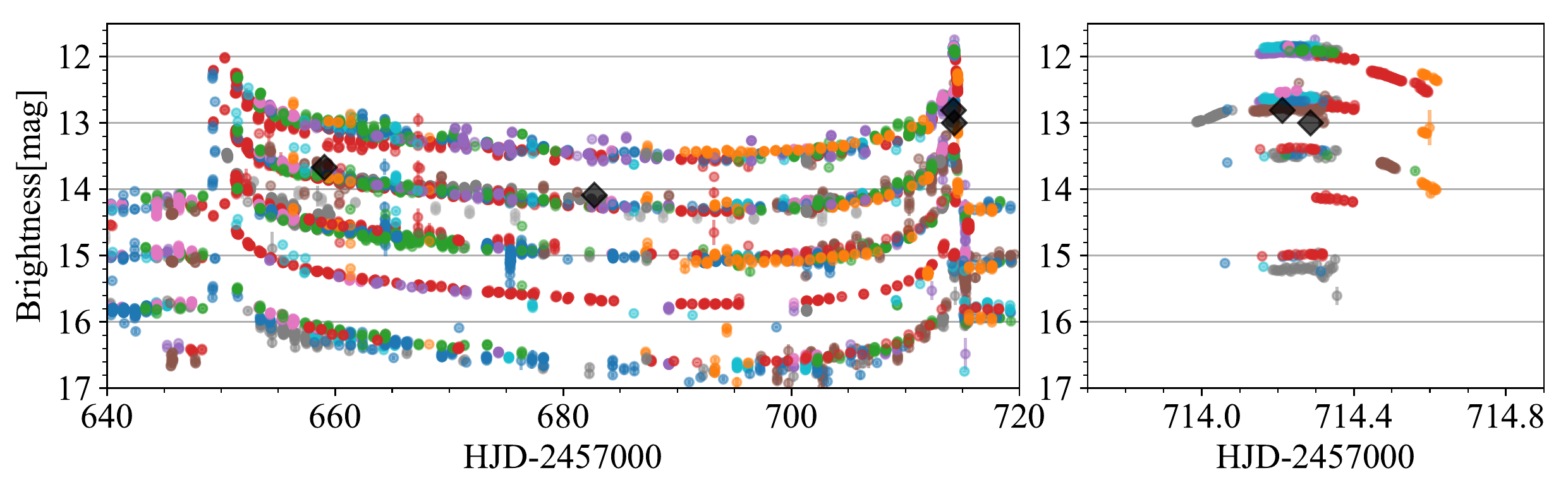}
\caption{Gaia, ASAS-SN, and follow-up photometric observations of Gaia16aye. 
Each observatory and observer is marked with a different colour. The marker is explained in the legend. 
The figure shows only the follow-up data, which were automatically calibrated using the Cambridge Photometric Calibration Server. 
The upper panel shows the entire event, and the bottom panels show a zoom on the second pair of caustic crossings (left) and a detail of the fourth caustic crossing (right).
}
\label{fig:lc}
\end{center}
\end{figure*}

\section{Discovery and follow-up of Gaia16aye}

%

Gaia16aye was found during the regular examination of the photometric data collected by the {\Gaia} mission. 
%
{\Gaia} is a space mission of the European Space Agency (ESA) in science operation since 2014. 
Its main goal is to collect high-precision astrometric data, that is, positions, proper motions, and parallaxes, of all stars on the sky down to about 20.7 mag in {\Gaia} G band \citep{GaiaPrusti, GaiaDR2photo}. 
While {\Gaia} scans the sky multiple times, it provides near-real-time photometric data, which can be used to detect unexpected changes in the brightness or appearance of new objects from all over the sky.
This is dealt with by the {\Gaia} Science Alerts system  \citep{Wyrzykowski2012, Hodgkin2013, Wyrzykowski2014}, which processes daily portions of the spacecraft data and produces alerts on potentially interesting transients. 
The main purpose of the publication of the alerts from Gaia is to enable the astronomical community to study the unexpected and temporary events. Photometric follow-up is necessary in particular in the case of microlensing events in order to fill the gaps between Gaia observations and subsequently construct a densely sampled light curve, sensitive to short-lived anomalies and deviations to the standard microlensing evolution  \citep[\eg][]{Wyrzykowski2012b}.

Gaia16aye was identified as an alert in the data chunk from 5 August 2016, processed on 8 August by the {\Gaia} Science Alerts pipeline ({\it AlertPipe}), and published on {\Gaia} Science Alerts webpages\footnote{\href{http://gsaweb.ast.cam.ac.uk/alerts/alert/Gaia16aye}{http://gsaweb.ast.cam.ac.uk/alerts/alert/Gaia16aye}} on 9 August 2016, 10:45 GMT.  The full {\Gaia} photometry of Gaia16aye is listed in Table \ref{tab:gaiaphot}. 

The alert was triggered by a significant change in brightness of an otherwise constant-brightness star with G=15.51 mag. 
The star has a counterpart in the 2MASS catalogue as 2MASS19400112+3007533 at RA,Dec (J2000.0) = 19:40:01.14, 30:07:53.36, and its source Id in {\Gaia} DR2 is 
2032454944878107008 \citep{GaiaDR2}.
Its Galactic coordinates are $l$,$b$ = 64.999872, 3.839052 deg, which locates Gaia16aye well in the northern part of the Galactic Plane towards the Cygnus constellation (see Fig. \ref{fig:finding}).

Gaia collected its first observation of this star in October 2014, and until the alert in August 2016, there were no significant brightness variation in its light curve. 
Additionally, this part of the sky was observed prior to {\Gaia} in 2011--2013 as part of a Nova Patrol  \citep{Sokolovsky2014}, and no previous brightenings were detected at a limiting magnitude of V$\approx$14.2. 

In the case of Gaia16aye the follow-up was initiated because the source at its baseline was relatively bright and easily accessible for a broad range of telescopes with smaller apertures. Moreover, microlensing events brighter than about G=16 mag will have Gaia astrometric data of sufficient accuracy in order to detect the astrometric microlensing signal \citep{Rybicki2018}. 
For this purpose, we have organised a network of volunteering telescopes and observers who respond to Gaia alerts, in particular to microlensing event candidates, and invest their observing time to provide dense coverage of the light curve. 
The network is arranged under the Time-Domain work package of the European Commission's Optical Infrared Coordination Network for Astronomy (OPTICON) grant\footnote{\href{https://www.astro-opticon.org/h2020/network/na4.html}{https://www.astro-opticon.org/h2020/network/na4.html}}.

The follow-up observations started immediately after the announcement of the alert (the list of telescopes and their acronyms is provided in Tab.\ref{tab:telescopesnames}), with the first data points taken on the night 9/10 Aug 2016 with the 0.6m Akdeniz Univ. UBT60 telescope in the TUBITAK National Observatory, Antalya, 
the SAI Southern Station in Crimea, 
the pt5m telescope at the Roque de los Muchachos Observatory on La Palma \citep{Hardy2015}, 
the 0.8m Telescopi Joan Oro (TJO) at l'Observatori Astronomic del Montsec,
and the 0.8m robotic APT2 telescope in Serra La Nave (Catania).
The data showed a curious evolution and a gradual rise (0.1 mag/day) in the light curve without change in colour, which is atypical for many known types of variable and cataclysmic variable stars. 
On the night 13/14 Aug 2016 (HJD$^{\prime} \equiv$ HJD-2450000.0 $\sim$ 7614.5) the object reached a peak V=13.8 mag (B-V=1.6 mag, I=12.2 mag), as detected by ATP2 and TJO, which was followed by a sudden drop by about 2 magnitudes. 
Alerted by the unusual shape of the light curve, we obtained spectra of Gaia16aye with the 1.22m Asiago telescope on 11 August and with the 2.0m Liverpool Telescope (LT, La Palma) on 12 August, which were consistent with a normal K8-M2 type star \citep{Bakis2016}.
The stellar spectra along with the shape of the light curve implied that Gaia16aye was a binary microlensing event, which was detected by {\Gaia} at its plateau between the two caustic crossings, and we have observed the caustic exit with clear signatures of the finite source effects. 

The continued follow-up after the first caustic exit revealed a very slow gradual rise in brightness (around 0.1 mag in a month). On 17 September 2016, it increased sharply by 2 mag (first spotted by the APT2 telescope), indicating the second caustic entry. The caustic crossing again showed a broad and long-lasting effect of finite source size (flattened peak), lasting for nearly 48 hours between HJD$^{\prime}$=7649.4 and 7651.4 and reaching about V=13.6 mag and I=12 mag. 
The caustic crossing was densely covered by the Liverpool Telescope and the 0.6m Ostrowik Observatory near Warsaw, Poland.  

Following the second caustic entry, the object remained very bright (I$\sim$12-14 mag) and was observed by multiple telescopes from around the globe, both photometrically and spectroscopically. 
The complete list of telescopes and instruments involved in the follow-up observations of Gaia16aye is shown in Table \ref{tab:telescopesnames}, and their parameters are gathered in Table \ref{tab:telescopes-instruments} in the appendix. 
In total, more than 25,000 photometric and more than 20 spectroscopic observations were taken over the period of about two years. 
In early November 2016, the brightness trend changed from falling to rising, as expected for binary events during the caustic crossing \citep{Nesci2016, Khamitov2016a}.
A simple preliminary model for the binary microlensing event predicted the caustic exit to occur around November 20.8 UT (HJD$^{\prime}$=7713.3) and the caustic crossing to last about seven hours \citep{Mroz2016}.
In order to catch and cover the caustic exit well, an intensive observing campaign was begun, involving also amateur astronomical associations (including the British Astronomical Association and the German Haus der Astronomie) and school pupils. 
The observations were also reported live on Twitter (hashtag {\it \#Gaia16aye}). 
A DDT observing time was allocated at the William Herschel Telescope (WHT/ACAM) and the Telescopio Nazionale Galileo (TNG/DOLORES) to provide low- and high-resolution spectroscopy at times close to the peak.
However, the actual peak occurred about 20 hours later than expected, on 21 November 16 UT (7714.17), and was followed by TRT-GAO, Aries130, CrAO, AUT25, T60, T100, RTT150 (detection of the fourth caustic was reported in \citealt{Khamitov2016b}), Montarrenti, Bialkow, Ostrowik, Krakow50, OndrejovD50, LT, pt5m, Salerno, and UCLO, 
spanning the whole globe, which provided 24-hour coverage of the caustic exit.
The sequence of spectroscopic observations before and at the very peak was taken with the IDS instrument on the Isaac Newton Telescope (INT).
After the peak at 11.85 mag in I band, the event brightness smoothly declined, as caught by Swarthmore24, DEMONEXT, and AAVSO.
The first datapoint taken on the next night from India (Aries130 telescope) showed I=14.33 mag, indicating the complete exit from the caustic. 
The event then again began to rise very slowly, with a rate of 1 mag over four months, and it exhibited a smooth peak on 5 May 2017 (HJD'=7878), reaching I=13.3 mag (G$\sim$14 mag)  \citep{Wyrzykowski2017}.
After this, the light curve declined slowly and reached the pre-alert level in November 2017, at G=15.5 mag. 
We continued our photometric follow-up for another year to confirm that there was no further re-brightening.
Throughout the event, the All-Sky Automated Survey for SuperNovae (ASAS-SN) \citep{Shappee2014, Kochanek2017ASASSN} observed Gaia16aye serendipitously with a typical cadence of between two and five days. 
Its data cover various parts of the light curve of the event, including the part before the {\Gaia} alert, where a smooth rise and the first caustic entry occurred. 

\subsection{Ground-based photometry calibrations}
Each observatory processed the raw data with their own standard data reduction procedures to create bias, dark-subtracted, and flat-fielded images. Then, the images were solved astrometrically, most often with the use of 
\texttt{Astrometry.net} code (\citealt{Hogg2008, Lang2010}),
and the instrumental photometry for all objects within the field of view was derived with a variety of tools, including Source EXtractor \citep{Sextractor} and Daophot \citep{Daophot}.
The lists of detected sources with their measured instrumental 
magnitudes were uploaded to the Cambridge Photometric Calibration Server (CPCS)\footnote{\href{http://gsaweb.ast.cam.ac.uk/followup}{http://gsaweb.ast.cam.ac.uk/followup}}, designed and maintained by Sergey Koposov and Lukasz Wyrzykowski.
The CPCS matches the field stars to a reference catalogue, identifies the 
target source, and determines which filter was used for observations. 
This tool acted as a central repository for all the data, but primarily, it standardised the data into a homogenous photometric system. It relied on available archival catalogues of this patch of the sky (primarily the AAVSO Photometric All-Sky Survey, APASS, and the Pan-STARRS1 Surveys, PS1) and derived zero-points for each of the observations. 
The use of a common repository allowed for near-real-time tracking of the evolution of the event, which is particularly important near the caustic entry and exit. Photometric data were uploaded by the observers within minutes of the observation, which facilitated detailed planning of the spectroscopic follow-up. 

The list of all the ground-based photometric observations is summarised in Table \ref{tab:followupphot-summary}
 and the photometric observations are listed in Table \ref{tab:followupphot} that is available in the appendix.
 The full table contains 23,730 entries and is available in the electronic version of the paper. 
 Figure \ref{fig:lc} shows all follow-up measurements collected for Gaia16aye over a period of about one and a half years.

\begin{table*}
\caption{Telescopes used in the photometric follow-up observations of Gaia16aye.}             
\begin{scriptsize}
\label{tab:telescopesnames}      
\centering                          
\begin{tabular}{l l l l l l}
\hline\hline                 
Telescope code & Telescope/observatory name & Location & Longitude [deg] & Latitude [deg] & Reference \\
\hline                        
AAVSO & American Association of Variable Star Observers & world-wide network, MA, USA & -- & -- & - \\ 
Akeno50 & 50-cm telescope, Akeno Observatory & Asao, Akeno-mura, Japan & 138.30 & 35.47 & - \\
APT2 & Automatic Photometric Telescope 2, & Serra La Nave, Mt. Etna, Italy  & 14.97 & 37.69 & - \\
 & Catania Astrophysical Observatory &  &  &  & \\
Aries130 & 1.30-m telescope, & Manora Peak, Nainital, India & 79.45 & 29.37 & - \\
 & Aryabhatta Research Institute of Observational Sciences & & & & \\ 
Aristarchos & Aristarchos Telescope, Helmos Observatory & Mt. Helmos, Peloponnese & 22.20 & 37.99 & \citet{Goudis2010} \\
ASASSN & All-Sky Automated Survey for Supernovae & world-wide network of 20 telescopes & -- & -- & \citet{Kochanek2017} \\
ASV1 & Astronomical Station Vidojevica 0.6 m & Vidojevica, near Prokuplje, Serbia & 21.56 & 43.14 & - \\
ASV2 & Astronomical Station Vidojevica 1.4 m & Vidojevica, near Prokuplje, Serbia & 21.56 & 43.14 & - \\
AUT25 & 25-cm telescope, Akdeniz University & Antalya, Turkey & 30.66 & 36.90 & - \\
BAS2 & Rozhen 2 m, National Astronomical Observatory,  & Rozhen, Bulgaria  & 24.74 & 41.70 & - \\
 & Bulgarian Academy of Sciences &  &  &  & \\
BAS50/70 & Schmidt-camera 50/70 cm, National Astronomical  & Rozhen, Bulgaria  & 24.74 & 41.70 & - \\
 & Observatory, Bulgarian Academy of Sciences &  &  &  & \\
Bialkow & Bia\l{}k\'ow Observatory,  & Bia\l{}k\'ow, Poland & 16.66 & 51.48 & - \\ 
 & Astronomical Institute of the University of Wroc\l{}aw &  &  &  &  \\
C2PU & C2PU-Omicron, & OCA, Calern Plateau, France & 6.92 & 43.75 & - \\
 & Center for Pedagogy in Planet and Universe sciences &  &  &  &  \\
Conti & Conti Private Observatory & MD, USA & -76.49 & 38.93 &  - \\
CrAO & Crimean Astrophysical Observatory & Nauchnyi, Crimea & 34.01 & 44.73 & - \\
DEMONEXT & DEdicated MONitor of EXotransits and Transients, & AZ, USA & -110.60 & 31.67 & \cite{Villanueva2018} \\ 
 & Winer Observatory &  & &  & \\ 
Foligno & Foligno Observatory & Perugia Province, Italy & 12.70 & 42.96 & - \\ 
HAO50 & Horten Astronomical Telescope & Nykirke, Horten, Norway & 10.39 & 59.43 & - \\ 
Krakow50 & 50-cm Cassegrain telescope, & Krak\'ow, Poland & 19.82 & 50.05 & - \\
 &  Astronomical Observatory of Jagiellonian University &  & &  & \\
Kryoneri & 1.2-m Kryoneri telescope, Kryoneri Observatory & Mt. Kyllini, Peloponnese, Greece & 22.63 & 38.07 & \citet{Xilouris2018}\\
LCO-Texas & Las Cumbres Observatory & McDonald Observatory, TX, USA & -104.02 & 30.67 & \citet{2013PASP..125.1031B}\\
LCO-Hawaii & Las Cumbres Observatory & Haleakala, HI, USA & -156.26 & 20.71 & \citet{2013PASP..125.1031B}\\
Leicester & University of Leicester Observatory & Oadby, UK & -1.07 & 52.61 & - \\ 
Loiano & 1.52 m Cassini Telescope, & INAF-Bologna, Loiano, Italy & 11.33 & 44.26 & - \\ 
 & INAF - Bologna Observatory of Astrophysics and Space Science & & &  &  \\ 
LOT1m & Lulin One-meter Telescope & Lulin Observatory, Taiwan & 120.87 & 23.47 & - \\
LT & Liverpool Telescope,  & La Palma, Spain & -17.88 & 28.76 & \citet{2004SPIE.5489..679S} \\
 & Roque de Los Muchachos Observatory &  &  &  &  \\
MAO165 & 1.65-m Ritchey–Chretien telescope,  & Mol\.etai, Kulionys, Lithuania & 25.56 & 55.32 & - \\
 &  Mol\.etai Astronomical Observatory &  &  &  &  \\
Mercator & Mercator Telescope,  & La Palma, Spain & -17.88 & 28.76 & - \\
 & Roque de Los Muchachos Observatory & &  &  & \\
Montarrenti & Montarrenti Observatory & Siena, Italy & 11.18 & 43.23 & - \\ 
OHP & T120, L'Observatoire de Haute-Provence & St. Michel, France & 5.71 & 43.93 & - \\ 
OndrejovD50 & D50 telescope, Astronomical Institute  & Ondrejov, Czech Rep. & 14.78 & 49.91 & - \\
 & of Academy of Sciences of the Czech Republic &  &  &  & \\
Ostrowik & Cassegrain telescope,  & Ostrowik, Poland & 21.42 & 52.09 & -\\
 &  Warsaw University Astronomical Observatory &  &  &  & \\
PIRATE & Physics Innovations Robotic Astronomical  & Tenerife, Spain & -16.51 & 28.30 & - \\
 &  Telescope Explorer Mark-III, Teide Observatory &  &  &  &  \citet{Kolb2018} \\
pt5m & 0.5m robotic telescope, & La Palma, Spain & -17.88 & 28.76 & \citet{Hardy2015} \\
 &  Roque de Los Muchachos Observatory &  &  &  &  \\
RTT150 & 1.5-m Russian-Turkish Telescope,  & Mt. Bakirlitepe, Antalya, Turkey & 30.33 & 36.83 & - \\
 &  TUBITAK National Observatory &  &  & &  \\
SAI & 60-cm Zeiss-2 telescope, Moscow State Univercity & Nauchnyi, Crimea & 34.01 & 44.73 & - \\
 &  observational station of Sternberg Astronomical Institute &  &  &  & \\
Salerno & Salerno University Observatory & Fisciano, Italy  & 14.79 & 40.78 & - \\
SKAS-KFU28 & C28 CGEM-1100 telescope,  & Zelenchukskaya, Caucasus, Russia & 41.43 & 43.65 & - \\
 &  Zelenchukskaya Station of Kazan Federal University &  &  &  &  \\
Skinakas & 1.3-m telescope, Skinakas Observatory & Skinakas, Crete, Greece & 24.90 & 35.21 & - \\ 
SKYNET & Skynet Robotic Telescope Network,  & WI, USA & -88.56 & 42.57 & - \\ 
 &  41-inch telescope, Yerkes Observatory &  &  &  &  \\ 
Swarthmore24 & 24-inch telescope, Peter van de Kamp Observatory & Swarthmore College, PA, USA & -75.36 & 39.91 & - \\
T60 & 60-cm telescope, TUBITAK National Observatory  & Mt. Bakirlitepe, Antalya, Turkey & 30.33 & 36.83 & - \\
T100 & 1.0-m telescope, TUBITAK National Observatory  & Mt. Bakirlitepe, Antalya, Turkey & 30.33 & 36.83 & - \\
TJO & Joan Or\'o Telescope, Montsec Observatory & Sant Esteve de la Sarga, Lleida, Spain & 0.73 & 42.03 & - \\
TRT-GAO & Thai Robotic Telescope GAO, Yunnan Observatory & Phoenix Mountain, Kunming, China & 105.03 & 26.70 & - \\
TRT-TNO & Thai Robotic Telescope TNO, & Doi Inthanon, Chiang Mai, Thailand & 98.48 & 18.57 & - \\
 &  Thai National Observatory &  &  &  & \\
UCLO-C14E & University College London Observatory, C14 East & Mill Hill, London, UK & -0.24 & 51.61 & - \\
UCLO-C14W & University College London Observatory, C14 West & Mill Hill, London, UK & -0.24 & 51.61 & - \\
UBT60 & Akdeniz University Telescope,  & Mt. Bakirlitepe, Antalya, Turkey & 30.33 & 36.83 & -\\
 &  TUBITAK National Observatory &  &  &  & \\
Watcher & 40-cm telescope, Boyden Observatory & Orange Free State, South Africa & 26.40 & -29.04 &  \cite{French2004} \\
WHT-ACAM & William Herschel Telescope,  & La Palma, Spain & -17.88 & 28.76 & - \\ 
 &  Roque de Los Muchachos Observatory &  &  &  &  \\ 
Wise1m & 1.0-m telescope, Wise Observatory & Mitzpe Ramon, Israel & 34.76 & 30.60 & - \\
WiseC28 & C28 Jay Baum Rich telescope, Wise Observatory & Mitzpe Ramon, Israel & 34.76 & 30.60 & - \\

\hline                                   
\end{tabular}
\end{scriptsize}
\end{table*}

\begin{table*}
\caption{Summary of observations taken by the observatories involved in the photometric follow-up of Gaia16aye. In brackets we list the best-matching filters as found by the Calibration Server. Asterisks mark data that were not uploaded to the CPCS.} 
\label{tab:followupphot-summary}      
\centering                          
\begin{tabular}{l l l l l}
\hline\hline                 
Telescope code  & First epoch & Last epoch  & Npoints (filter), Npoints (filter2), etc. \\
 & [HJD$-$2450000] & [HJD$-$2450000] & \\
\hline                        
AAVSO & 7653.283 & 7714.561 & 288(V) 151(i) 95(r)  \\
Akeno50 & 7711.012 & 7715.301 & 169(r)$^*$ \\
APT2 &  7612.294 & 8055.256 & 285(B) 467(V) 439(i) 452(r)  \\
Aries130  & 7714.070 & 7718.030 & 6(B) 6(V) 6(R) 6(I)  \\
Aristarchos  & 8035.219 & 8039.086 & 2(B) 2(V) 1(g) 6(i) 44(r)  \\
ASASSN & 7547.097 & 7907.897 & 68(V)$^*$ \\
ASV1 & 7929.570 & 8079.302 & 11(B) 34(V) 36(i) 28(r)  \\
ASV2 & 7628.483 & 7924.511 & 42(B) 64(V) 1(g) 69(i) 73(r)  \\
AUT25  & 7712.258 & 7715.274 & 136(i) 142(r)  \\
BAS2+BAS50/70 &  7687.225 & 7933.497 & 8(B) 23(V) 9(g) 28(i) 31(r)  \\
Bialkow &  7619.340 & 8028.296 & 218(B) 499(V) 657(i) 641(r)  \\
C2PU  & 7637.331 & 7878.619 & 8(V) 41(r)  \\
Conti & 7714.470  &  7714.510 & 38(V) \\
CrAO &  7710.306 & 7871.562 & 639(r)  \\
DEMONEXT &  7690.672 & 8162.029 & 476(V) 483(i) 427(r)  \\
Foligno &  7654.361 & 7719.251 & 11(V)  \\
HAO50 & 7818.318 & 8056.320 & 22(V)$^*$, 10(R)$^*$ \\
Krakow50  & 7659.243 & 7919.552 & 17(B) 44(V) 49(i) 60(r)  \\
Kryoneri  & 7652.327 & 8039.210 & 92(i) 96(r)  \\
LCO-Texas &  7663.570 & 7904.530 & 63(B) 70(V) 30(g) 29(i) 94(r) \\
LCO-Hawaii &  6792.778 & 7708.778 & 197(gp)$^*$, 318(rp)$^*$, 518(ip)$^*$, 294(V)$^*$, 146(B)$^*$, 24(R)$^*$, 12(I)$^*$\\
Leicester & 7645.461 & 8063.274 & 10(B) 9(V) 3(i) 1(r)  \\
Loiano &  7660.301 & 7709.269 & 77(B) 66(V) 108(g) 119(i) 164(r)  \\
LOT1m &  7711.936 & 7888.223 & 54(g) 59(i) 55(r)  \\
LT &  7647.327 & 7976.490 & 2(V) 362(g) 415(i) 488(r)  \\
MAO165& 7680.350 & 7997.400 & 6(B)$^*$ 31(V)$^*$ 34(R)$^*$ 27(I)$^*$ \\
Mercator &  7651.332 & 7657.397 & 7(g) 5(r)  \\
Montarrenti   & 7654.280 & 7929.545 & 92(r)  \\
OHP &  7665.329 & 8019.350 & 6(V) 3(g) 11(i) 13(r)  \\
OndrejovD50  & 7614.564 & 8095.253 & 397(B) 410(V) 413(i) 423(r)  \\
Ostrowik & 7619.303 & 7735.192 & 3(B) 42(V) 1(g) 185(i) 193(r)  \\
PIRATE & 7650.498 & 7849.748 & 1473(r) 713(V)   \\
pt5m & 7610.408 & 8094.350 & 205(B) 2452(V) 243(i) 266(r)  \\
RTT150 &  7657.696 & 7937.559 & 114(B) 112(V) 1(g) 1(i) 1(r)  \\
SAI & 7610.282 & 7613.265 & 16(B) 16(V) 18(r)  \\
Salerno & 7651.308 & 7765.244 & 610(R)$^*$ \\
SKAS-KFU28 & 7662.357 & 7846.548 & 124(B)$^*$ 158(G)$^*$ 170(R)$^*$ \\
Skinakas  & 7668.246 & 7993.770 & 5(B) 1(G) 5(V) 2(g) 6(i) 5(r)  \\
SKYNET & 7670.521 & 7729.487 & 6(g) 64(i) 38(r)  \\
Swarthmore24 & 7714.444 & 7954.598 & 287(i)  \\
T60 & 7670.862 & 8436.268 & 1(B) 9(V) 8(r) 8(i) \\
T100 & 7637.476 & 7963.499 & 27(B) 34(V) 24(g) 21(i) 21(r)  \\
TJO & 7610.503 & 8090.273 & 485(B) 563(V) 1(g) 494(i) 524(r) 2(z)  \\
TRT-GAO &  7712.986 & 7886.388 & 3(V) 1016(r)  \\
TRT-TNO &  7833.368 & 7843.437 & 41(i) 48(r)  \\
UCLO-C14E &  7678.287 & 7711.319 & 5(V) 28(r)  \\
UCLO-C14W &  7666.399 & 7955.577 & 122(i) 44(r)  \\
UBT60 &  7610.246 & 7715.274 & 279(B) 349(V) 440(i) 448(r)  \\
Watcher & 7617.004 & 8017.002 & 258(V) 264(i) 261(r)  \\
WHT-ACAM  & 7701.314 & 7701.375 & 26(g) 30(i) 30(r)  \\
Wise1m &7654.236 & 7749.173 & 305(i)  \\
WiseC28 &  7652.396 & 7660.294 & 25(i)  \\
\hline                                   
\end{tabular}
\end{table*}

\subsection{Gaia data}
Since October 2014 {\Gaia} collected 27 observations before the alert on the 5 August 2016.
In total, {\Gaia} observed Gaia16aye 84 times as of November 2018.
The G-band photometric data points collected by {\Gaia} are listed in Table \ref{tab:gaiaphot}. 
Photometric uncertainties are not provided for {\Gaia} alerts, and for this event we assumed 0.01 mag \citep{GaiaPrusti}, but as we show below, these were scaled to about 0.015 mag by requiring the microlensing model $\chi^2$ per degree of freedom to be 1.0.
Details of the {\Gaia} photometric system and its calibrations can be found in \cite{GaiaDR2photo}. 

\begin{figure}
\begin{center}
\includegraphics[width=\columnwidth]{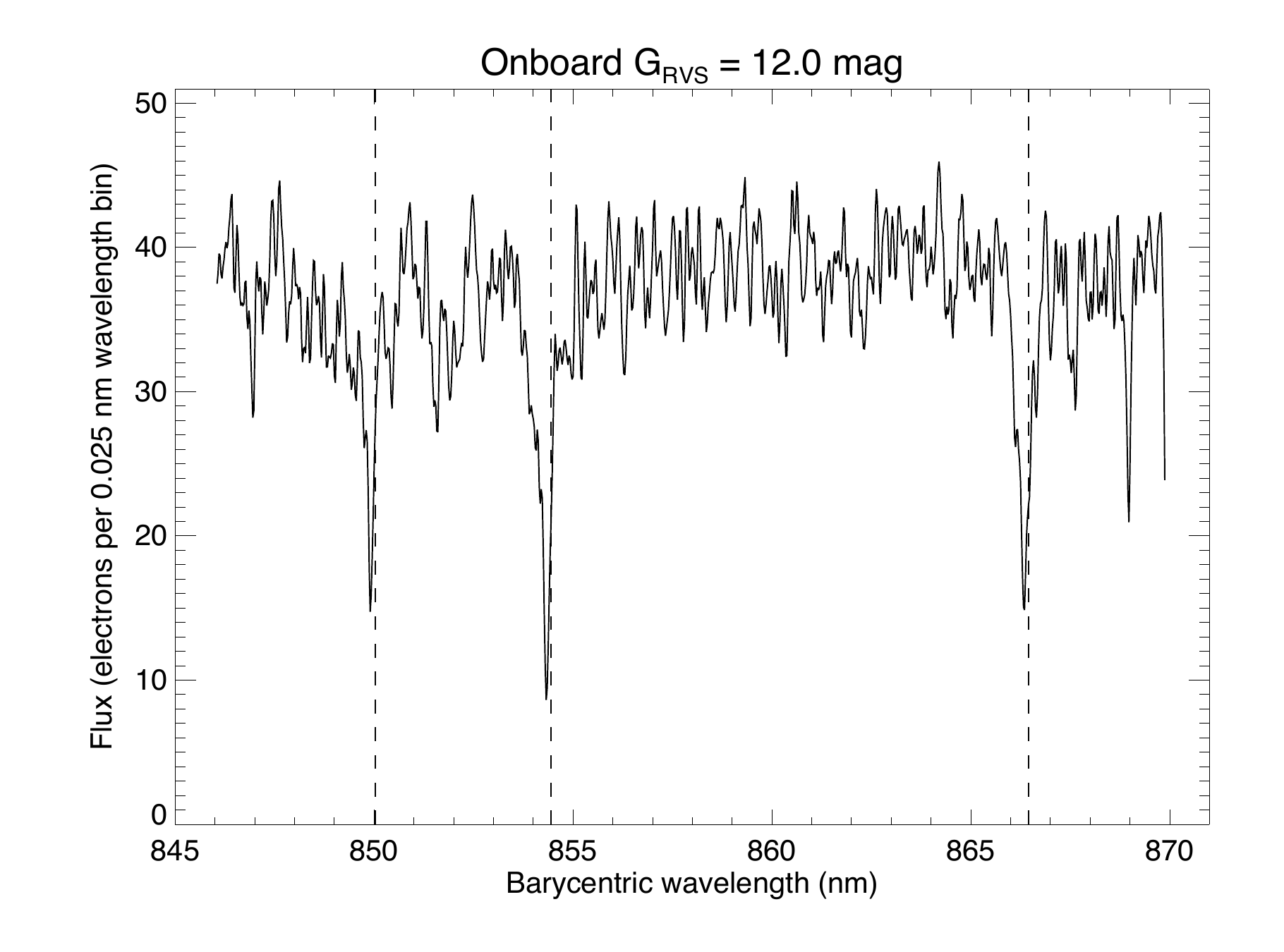}
\caption{Medium-resolution spectrum of the Gaia16aye event obtained with the {\Gaia} RVS at the brightest moment of the event as seen by {\Gaia} at the fourth caustic crossing. The CaII lines of the lensed source are clearly visible.}
\label{fig:rvs}
\end{center}
\end{figure}

{\it \textup{The}} on-board Radial Velocity Spectrometer (RVS) of \textit{Gaia}, 
{\bf
collects medium-resolution (R$\sim$11,700) spectra 
}
over the wavelength range 845-872 nm centred on the Calcium II triplet region of objects brighter than V$\sim$17 mag \citep{GaiaPrusti, GaiaRVS}.
However, individual spectra for selected observations are made available already for brighter {\Gaia} alerts using parts of the RVS data processing pipeline \citep{Sartoretti2018}.
For Gaia16aye the RVS collected a spectrum on 21 November 2016, 17:05:47 UT (HJD=2457714.21), see Figure \ref{fig:rvs}, the moment is caught by {\Gaia} at very high magnification, when Gaia16aye reached G=12.91 mag. The exposure time for the combined three RVS CCDs was 3$\times$4.4 seconds.

\begin{figure}
\begin{center}
\includegraphics[width=0.92\columnwidth]{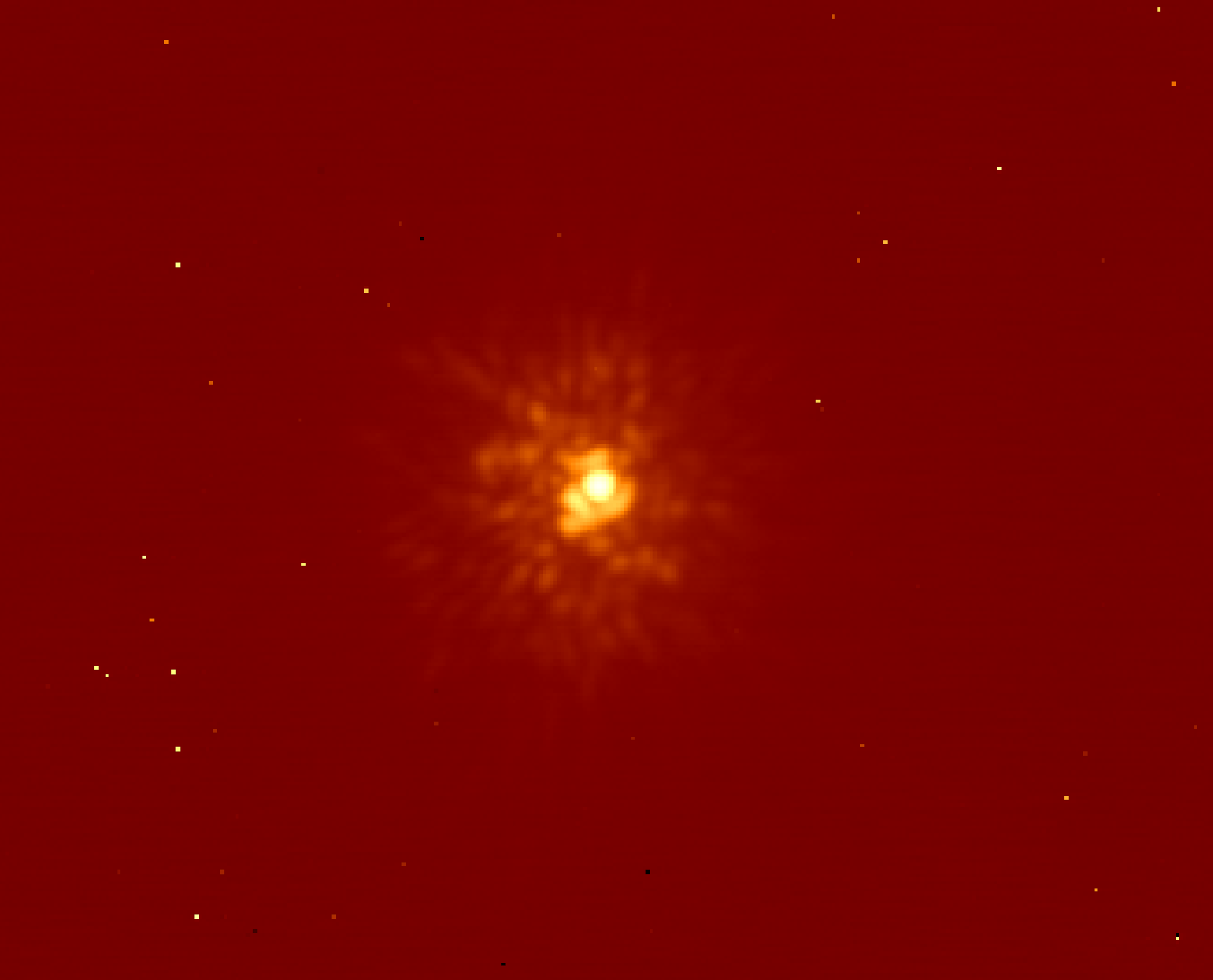}
\caption{Keck Adaptive Optics image of Gaia16aye taken between the third and fourth caustic crossing. The single star has an FWHM of about 52 mas. No other light sources contribute significantly to the blending in the event.}
\label{fig:keckao}
\end{center}
\end{figure}

\begin{figure}
\begin{center}
\includegraphics[width=\columnwidth]{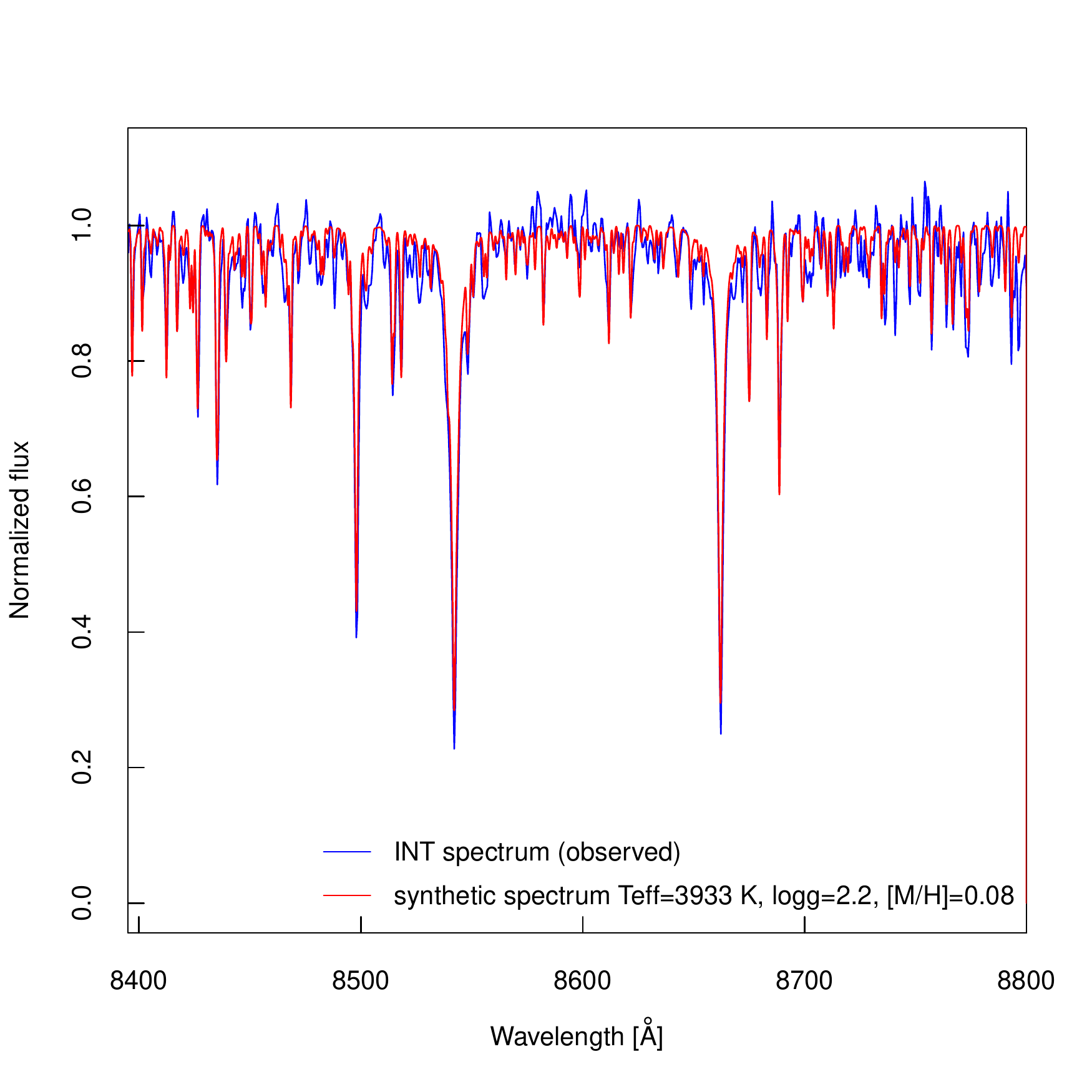}
\caption{Spectrum of the source of the Gaia16aye event ({\it blue}) taken using the 2.5 m INT/IDS on 19 November 2016 in comparison with a synthetic spectrum ({\it red}) calculated for the best-fit atmospheric parameters. The plot shows the Ca~II triplet region, $8400-8800$~\AA.}
\label{fig:spec-sourceCaII}
\end{center}
\end{figure}

\subsection{Spectroscopy}
Spectroscopic measurements of the event were obtained at various stages of its evolution. The list of spectroscopic observations is presented in Table \ref{tab:spectra}.
The very first set of spectra was taken with the Asiago 1.22~m telescope equipped with the DU440A-BU2 instrument, the Asiago 1.82~m telescope with AFOSC, and the SPRAT instrument on the 2~m Liverpool Telescope (LT), which showed no obvious features seen in outbursting Galactic variables. 
Other spectra gathered by the 5~m P200 Palomar Hale Telescope and by ACAM on the 4.2~m William Herschel Telescope (WHT) confirmed this behaviour.
This therefore led us to conclude that this is a microlensing event. 

We did not find significant differences between spectra taken at various consecutive stages of the event evolution. The features and general shape of the spectra were the same, regardless of whether the spectrum was recorded during amplification or in the baseline. This allows us to conclude that the spectra were dominated by radiation from the source, and contribution from the lens was negligible. 

Most of the spectra were obtained in low-resolution mode ($R\le1000$) and relatively poor weather conditions, which were useful for an early classification of the transient as a microlensing event. 
A more detailed analysis of the low-resolution spectra will be presented elsewhere (Zielinski M. et al., in prep.)

We also obtained spectra of higher resolution ($R\sim6500$) with the 2.5~m INT, La Palma, Canary Islands, during three consecutive nights on $19-21$ November 2016. 
{\bf
The INT spectra were obtained using
}
 the Intermediate Dispersion Spectrograph (IDS, Cassegrain Focal Station, 235 mm focal length camera RED+2) with the grating set to R1200Y, and a dispersion of 0.53\AA~pixel$^{-1}$ with a slit width projected onto the sky equal to 1.298$\arcsec$ (see Tab.~\ref{tab:spectra}, spectrum INT~3--5). The exposure time was 400~s for each spectrum centred at wavelength 8100\AA.

The spectra were processed by the observers with their own pipelines or in a standard way using IRAF\footnote{IRAF is distributed by the National Optical Astronomy Observatories, which are operated by the Association of Universities for Research in Astronomy, Inc., under cooperative agreement with the National Science Foundation.} tasks and scripts. The reduction procedure consisted of the usual bias- and dark-subtraction, flat-field correction, and wavelength calibration.

\begin{table*}
\caption{Summary of the spectroscopic observations of Gaia16aye.} 
\label{tab:spectra}      
\centering
\begin{tabular}{lccl}
\hline
Spectrum  & Observation date  & Wavelength range &  Telescope -- Instrument \\
ID        & HJD      & (\AA)          &         \\
\hline

LT 1  & 2457612.900668 & 4200 -- 7994 &  Liverpool Telescope -- SPRAT  \\
LT 2  & 2457617.940097 & 4200 -- 7994 &  Liverpool Telescope -- SPRAT \\
LT 3  & 2457643.845837 & 4200 -- 7994 &  Liverpool Telescope -- SPRAT  \\
\hline

WHT 1 & 2457701.3045827 & 4303 -- 9500 &  William Herschel Telescope -- ACAM \\
\hline

Palomar 1 &2457662.1047682 & 3100 -- 10200 &  Palomar Hale Telescope -- DBSP \\
Palomar 2 &2457932.6881373  & 3800 -- 10000 &  Palomar Hale Telescope -- DBSP \\
\hline

INT 1 & 2457703.4230518 &7550 -- 9000 &  Isaac Newton Telescope -- IDS; R831R grating \\
INT 2 & 2457706.3547417 & 7550 -- 9000 &  Isaac Newton Telescope -- IDS; R831R grating \\
INT 3 &2457712.2970278 &  7500 -- 8795 &  Isaac Newton Telescope -- IDS; R1200Y grating \\
INT 4 &2457713.2967616 & 7500 -- 8795  &  Isaac Newton Telescope -- IDS; R1200Y grating \\
INT 5 & 2457714.2949097 & 7500 -- 8795  &  Isaac Newton Telescope -- IDS; R1200Y grating \\
\hline

Asiago 1     &2457612.430953 &  3320 -- 7880  &  1.22m Reflector -- DU440A-BU2 \\
Asiago 2     &2457623.364186 &  4160 -- 6530  &  1.82m Reflector -- AFOSC; GR07 grating \\
Asiago 3a    &2457700.264730 & 8200 -- 9210   &  1.82m Reflector -- AFOSC;  VPH5 grating \\
Asiago 3b    &2457700.275567  & 5000 -- 9280  &  1.82m Reflector -- AFOSC;  VPH6 grating \\
Asiago 4a    &2457700.260113 &  8200 -- 9210  &  1.82m Reflector -- AFOSC;  VPH5 grating \\
Asiago 4b &  2457700.270951 &   5000 -- 9280   & 1.82m Reflector -- AFOSC;  VPH6 grating \\
Asiago 5a  &  2457722.263836 &  8200 -- 9210   & 1.82m Reflector -- AFOSC;  VPH5 grating \\
Asiago 5b  &  2457722.235417 & 5000 -- 9280   &  1.82m Reflector -- AFOSC;  VPH6 grating \\
Asiago 6a   &  2457723.246689 &  8200 -- 9210 &  1.82m Reflector -- AFOSC;  VPH5 grating \\
Asiago 6b   &  2457723.204078 & 5000 -- 9280  &  1.82m Reflector -- AFOSC;  VPH6 grating \\
\hline
\end{tabular}
\end{table*}

\subsection{Swift observations}
In order to rule out the possibility that Gaia16aye is some type of cataclysmic variable star outburst, we requested X-ray and ultraviolet Swift observations. 
Swift observed Gaia16aye for 1.5ks on 18 August 2016. Swift/XRT detected no X-ray source at the position of the transient with an upper limit of 0.0007$\pm$0.0007 cts/s (a single background photon appeared in the source region during the exposure).
Assuming a power-law emission with a photon index of 2 and
HI column density of $43.10\times 10^{20}$ cm$^{-2}$ (corresponding to the total Galactic column density in this direction \citep{Kalberla2005}), this translates into an unabsorbed 0.3-10~keV flux limit of $5.4\times 10^{-14}$ ergs/cm$^2$/s.

No ultraviolet source was detected by the UVOT instrument at the position of the transient. The upper limit at epoch HJD'=7618.86 was derived as $>$20.28 mag for UVM2-band (Vega system).

\subsection{Keck adaptive optics imaging}
The event was observed with Keck adaptive optics (AO) imaging on 8 October 2016 (HJD'=7669.7).
Figure \ref{fig:keckao} shows the 10 arcsec field of view obtained with the Keck AO instrument. 
The full width at half-maximum (FWHM) of the star is about 52 mas. The image shows a single object with no additional light sources in its neighbourhood. 
This indicates that no additional luminous components contributed to the observed light.

\section{Spectroscopy of the source star}
\label{sec:spectra}
During a microlensing event, the variation in the amplification changes the ratio of the flux from the source, while the blend or lens light remains at the same level. 
Therefore, the spectroscopic data obtained at different amplifications can be used to de-blend the light of the source from any additional constant components and to derive the source properties. 

In order to obtain the spectral type and stellar parameters of the Gaia16aye source, we used three spectra gathered by the 2.5~m INT.
Based on these spectra we were able to determine the atmospheric parameters of the microlensing source. We used a dedicated spectral analysis framework,  iSpec\footnote{\href{https://www.blancocuaresma.com/s/iSpec}{https://www.blancocuaresma.com/s/iSpec}} , which integrates several radiative transfer codes 
\citep{Blanco2014}. In our case, the SPECTRUM code was used \citep{Gray1994}, together with well-known Kurucz model atmospheres \citep{Kurucz1993} and solar abundances of chemical elements taken from \citet{Asplund2009}. 
The list of absorption lines with atomic data was taken from the VALD database 
\citep{Kupka2011}. We modelled synthetic spectra for the whole wavelength region between 7200--8800~\AA. The spectrum that was synthesized to the observational data with the lowest $\chi^{2}$ value constituted the final fit generated for specific atmospheric parameters: effective temperature ($T_{\rm{eff}}$), surface gravity ($\log~g$), and metallicity ([M/H]). For simplification purposes, we adopted solar values of micro-- and macroturbulence velocities and also neglected stellar rotation. The resolution of the synthetic spectra was fixed as $R=10\,000$. We applied this method to all three INT spectra independently and then averaged the results. The mean values for the source parameter in Gaia16aye were as follows: $T_{\rm{eff}}=3933\pm135$~K, $\log~g=2.20\pm1.44,$ and [M/H]$=0.08\pm0.41$~dex. 
Figure~\ref{fig:spec-sourceCaII} presents the best fit of the synthetic to observational INT spectrum in the same spectral region as was covered by the RVS spectrum of Gaia16aye, that is, 8400--8800~\AA ~(Ca~II triplet), generated for averaged parameter results. 
These parameters imply that the microlensing source is a K5-type giant or a super-giant with solar metallicity. 
We discuss the estimate for the source distance in the next section because it is first necessary to de-blend the light of the lens and the source, which is possible in the microlensing model.
We note that the asymmetry of the Gaia RVS lines is not visible in the same-resolution INT/IDS spectrum, and we suspect that the broadening visible in the Gaia spectrum is a result of a stack of spectra from separate RVS CCDs.

\section{Microlensing model}

\subsection{Data preparation}

The data sets we used in the modelling are listed in Table \ref{tab:followupphot-model} in the appendix.
Because the microlensing model is complex, we had to restrict the number of data points that were used.
We chose data sets that cover large parts of the light curve or important features (such as caustics).
Some of the available data sets were also disregarded because they showed strong systematic variations in residuals from the best-fit model, which are not supported by other data sets. We used observations collected in the Cousins $I$ or Sloan $i$ band because the signal-to-noise ratio in these filters is highest. The only exceptions were {\Gaia} ($G$-band filter) and ASAS-SN data ($V$ band), which cover large portions of the light curve, especially before the transient alert.

Calculating microlensing magnifications (especially during caustic crossings) requires much computational time. We thus binned the data to speed up the modelling. We commonly used one-day bins, except for caustic crossings (when brightness variations during one night are substantial), for which we used 0.5 hr or 1 hr bins. {\Gaia} and ASAS-SN data were not binned. 

We rescaled the error bars, so that $\chi^2/\mathrm{dof}\sim1$ for each data set. The error bars were corrected using the formula $\sigma_{i,\mathrm{new}}=\sqrt{(\gamma\sigma_i)^2+\epsilon^2}$. Coefficients $\gamma$ and $\epsilon$ for each data set are shown in Table~\ref{tab:modelling}. The final light curve is presented in Fig. \ref{fig:lc1}.

\subsection{Binary lens model}
The simplest model describing a microlensing event caused by a binary system needs seven parameters: the time of the closest approach between the source and the centre of mass of the lens $t_0$, the projected separation between source and barycenter of the lens at that time $u_0$ (in Einstein radius units), the Einstein crossing time $t_{\rm E}$, the mass ratio of the lens components $q$, the projected separation between two binary components $s$, the angle between the source-lens relative trajectory and the binary axis $\alpha$, and the angular radius of the source $\rho$ normalised to the Einstein radius (Eq.\ref{eq:thetaE}). 

\begin{figure}
\centering
\includegraphics[width=0.95\columnwidth]{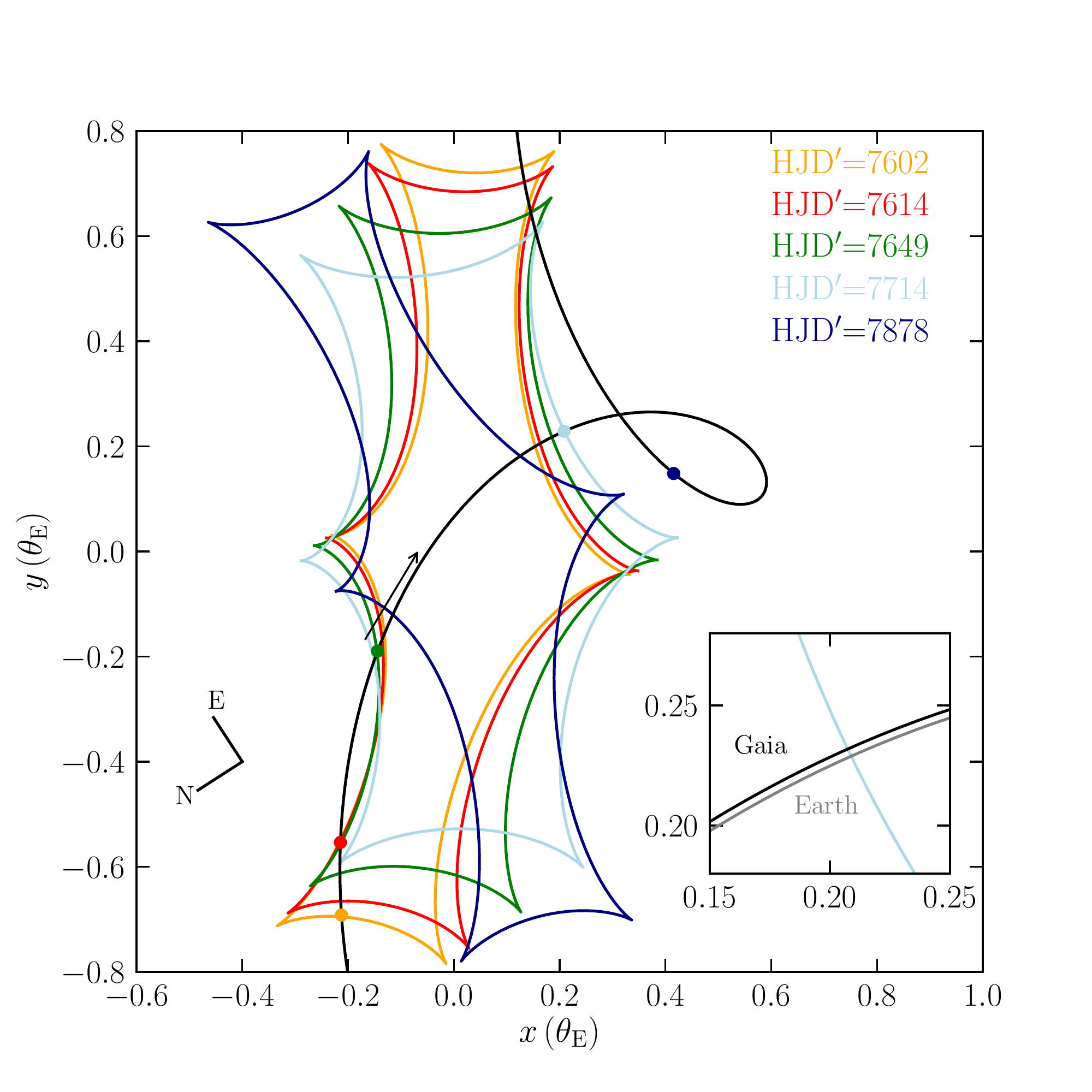}
\caption{Caustic curves corresponding to the best-fitting model of Gaia16aye. The lens-source relative trajectory is shown by a black curve. The barycenter of the lens is at $(0,0)$ and the lens components are located along the $x$ -axis at time $t_{\rm 0,kep}=7675$. Caustics are plotted at the times of caustic crossings; the large points are marked with respective colours. The inset shows a zoom on the trajectory of the Earth and Gaia at the moment of the caustic crossing around $\mathrm{HJD}'\sim7714$. }
\label{fig:caustic}
\end{figure}

This simple model is insufficient to explain all features in the light curve. We therefore included additional parameters that describe second-order effects: the orbital motion of the Earth (microlensing parallax) and the orbital motion of the lens. The microlensing parallax $\boldsymbol{\pi}_{\rm E}=(\pi_{\rm E,N},\pi_{\rm E,E})$ is a vector quantity:
$
\boldsymbol{\pi}_{\rm E}= \frac{\pi_{\rm rel}}{\theta_{\rm E}}\frac{\boldsymbol{\mu}_{\rm rel}}{\mu_{\rm rel}},
$
where $\boldsymbol{\mu}_{\rm rel}$ is the relative lens-source proper motion \citep{Gould2000b}. It describes the shape of the relative lens-source trajectory (Fig. \ref{fig:caustic}). The microlensing parallax can also be measured using simultaneous observations from two separated observatories, for exmaple, from the ground and a distant satellite \citep{Refsdal1966,Gould1994}. Because {\Gaia} is located at the $L_2$ Lagrange point (about 0.01\,au from the Earth) and the Einstein radius projected onto the observer's plane is $\mathrm{au}/\pi_{\rm E}\approx 2.5$\,au, 
the magnification gradient changes by less than the data precision throughout most of the light curve (see Fig. \ref{fig:lc2}). 
Fortunately, two \textit{Gaia} measurements were collected near $\mathrm{HJD}'\sim 7714$, when the space-parallax signal is strongest due to rapid change in magnification near the caustic.
Therefore, we included the space-parallax and {\Gaia} observations in the final modelling.

The orbital motion of the lens can in the simplest scenario be approximated as linear changes of separation $s(t)=s_0 + \dot{s}(t-t_{\rm 0,kep})$ and angle $\alpha(t)=\alpha_0 + \dot{\alpha}(t-t_{\rm 0,kep})$, $t_{\rm 0,kep}$ can be any arbitrary moment of time and is not a fit parameter \citep{Albrow2000}. 
This approximation, which works well for the majority of binary microlensing events, is insufficient in this case.

We have to describe the orbital motion of the lens using a full Keplerian approach \citep{Skowron2011}. This model is parameterised by the physical relative 3D position and velocity of the secondary component relative to the primary,
$
\Delta\boldsymbol{r} = D_l \theta_{\rm E} (s_0,0,s_z), \Delta\boldsymbol{v} = D_l\theta_{\rm E}s_0(\gamma_x,\gamma_y,\gamma_z)
$
at time $t_{\rm 0,kep}$. For a given angular radius of the source star $\theta_*$ and source distance $D_s$, we can calculate the angular Einstein radius $\theta_{\rm E} = \theta_* / \rho$ and distance to the lens $D_l = \mathrm{au} / (\theta_{\rm E}\pi_{\rm E}+\mathrm{au}/D_s)$. Subsequently, positions and velocities can be transformed to orbital elements of the binary (semi-major axis $a$, orbital period $P$, eccentricity $e$, inclination $i$, longitude of the ascending node $\Omega$, argument of periapsis $\omega$, and time of periastron $t_{\rm peri}$). 
These can be used to calculate the projected position of both components on the sky at any moment in time.

In all previous cases of binary events with significant binary motion, Keplerian orbital motion provided only a small improvement relative to the linear approximation \citep{Skowron2011,shin2012}. This is not the case here, because, as we show below, the orbital period of the lens is similar to the duration of the event \citep[\eg][]{Penny2011}. Modelling of this event is an iterative process: for given microlensing parameters, we estimated the angular radius and distance to the source, we calculated best-fit microlensing parameters, and we repeated the procedure until all parameters converged.

The best-fit microlensing parameters are presented in Table \ref{tab:params}. 
Uncertainties were calculated using the Markov Chain Monte Carlo approach (MCMC) \citep{foreman2013} and represent 68\% confidence intervals of marginalized posterior distributions. 
We note that another degenerate solution exists for the microlensing model that differs only by the signs of $s_z$ and $\gamma_z$ ($(s_z,\gamma_z)\rightarrow-(s_z,\gamma_z)$). 
The second solution has the same physical parameters (except for $\Omega\rightarrow\pi-\Omega$ and $\omega\rightarrow\omega-\pi$) and differs by the sign of the radial velocity. Thus, the degeneracy can be broken with additional radial velocity measurements of the lens \citep{Skowron2011}.

\subsection{Source star}
Spectroscopic observations of the event indicate that the source is a K5-type giant or a super-giant. If the effective temperature of the source were higher than 4250\,K, TiO absorption features would be invisible. If the temperature were lower than 3800\,K, these features would be stronger than those in the observed spectra. Spectral modelling indicates that the effective temperature of the source is $3933 \pm 135$\,K. According to \citet{hou}, the intrinsic Johnson-Cousins colours of a star of this spectral type and solar metallicity should be $(V-R)_0=0.83 ^{+0.03}_{-0.12}$, $(V-I)_0=1.60^{+0.03}_{-0.12}$ and $(V-K)_0=3.64^{+0.11}_{-0.37}$ (error bars correspond to the source of K4- and M0-type, respectively).

We used a model-independent regression to calculate the observed colours of the source (we used observations collected in the Bialkow Observatory, which were calibrated to the standard system): $V-R=0.99 \pm 0.01$ and $V-I=1.91 \pm 0.01$. Thus, the colour excess is $E(V-I)=0.31$ and $E(V-R)=0.16$, consistent with the standard reddening law \citep{Cardelli1989} and $A_V=0.62$. 

According to the best-fitting microlensing model, the amount of light coming from the magnified source is 
$V_{\rm s}=16.61 \pm 0.02$ and $I_{\rm s}=14.70 \pm 0.02$. 
The $V$-band brightness of the source after correcting for extinction is therefore $V_0 = 15.99$ mag. 
Subsequently, we used the colour--surface brightness relations for giants from \citet{adams2018} to estimate the angular radius of the source: $\theta_* = 9.2 \pm 0.7\,\mu$as. Because the linear radius of giants of this spectral type is about $31\pm 6\,R_{\odot}$ \citep{dyck1996}, the source is located about $15.7 \pm 3.0$\,kpc from the Sun, but the uncertainties are large. For the modelling we assumed $D_s=15$\,kpc. We note that the exact value of the distance has in practice a very small effect on the final models because $\pi_s \ll \theta_{\rm E}\pi_{\rm E}$.

\subsection{Physical parameters of the binary lens}
\begin{figure*}
\centering
\includegraphics[width=\textwidth]{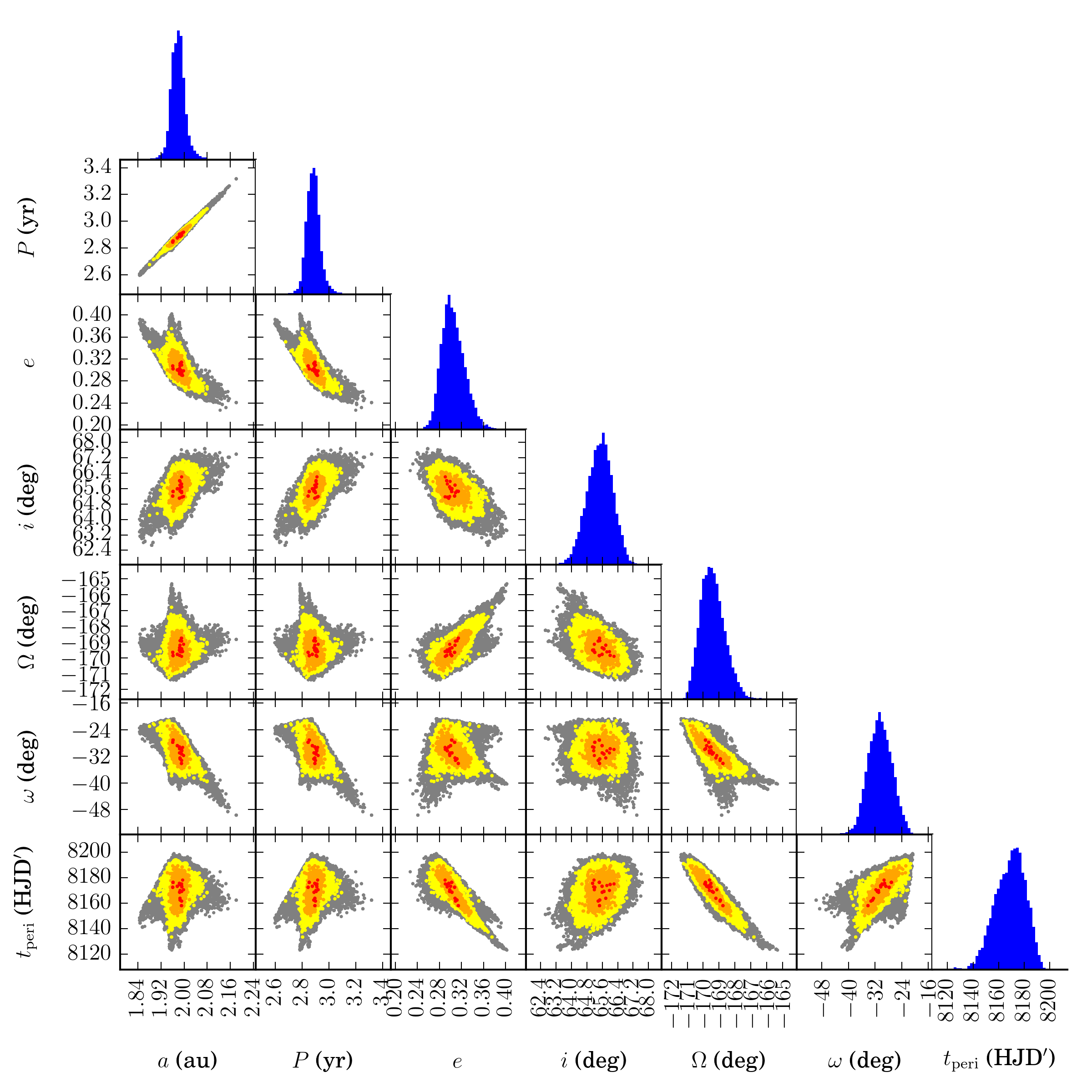}
\caption{Orbital elements of Gaia16aye. The panels show 2D and 1D projections of posterior distributions in the space of Kepler parameters. Red, orange, and yellow points mark $1\sigma$, $2\sigma$, and $3\sigma$ confidence regions, respectively.}
\label{fig:orbital}
\end{figure*}

The Gaia16aye microlensing model allows us to convert microlensing quantities into physical properties of the lensing binary system. Finite source effects over the caustics enabled us to measure the angular Einstein radius,
$$
\theta_{\rm E}=\frac{\theta_*}{\rho}=3.04 \pm 0.24\,\mathrm{mas}
$$
and the relative lens-source proper motion,
$$
\mu_{\rm rel} = \frac{\theta_{\rm E}}{t_{\rm E}}=10.1 \pm 0.8 \,\mathrm{mas\,yr}^{-1}.
$$
Because the microlensing parallax was precisely measured from the light curve (Table~\ref{tab:params}), we were able to measure the total mass of the lens,
$$
M = \frac{\theta_{\rm E}}{\kappa\pi_{\rm E}}=0.93 \pm 0.09 M_{\odot}
$$
and its distance,
$$
D_l = \frac{\mathrm{au}}{\theta_{\rm E}\pi_{\rm E}+\mathrm{au}/D_s}=780 \pm 60\,\mathrm{pc}.
$$
The orbital parameters of the lens were calculated using the prescriptions from \citet{Skowron2011} based on the full information about the relative 3D position and velocity of the secondary star relative to the primary. All physical parameters of the lens are given in Table~\ref{tab:physical}.
Figure \ref{fig:orbital} shows the orbital parameters and their confidence ranges as derived from the MCMC sampling of the microlensing model. 
Our microlensing model also allowed us to separate the flux from the source and the unmagnified blended flux (that comes from the lens, as we show below): $V_{\rm blend}=17.98 \pm 0.02$, $R_{\rm blend}=17.05 \pm 0.02$, and $I_{\rm blend}=16.09 \pm 0.02$ (Table~\ref{tab:params}).

\begin{table}
\caption{Data sets used in the modelling}
\label{tab:modelling}
\centering
\begin{tabular}{lrrrr}
\hline \hline
Observatory & Filter & Number & $\gamma$ & $\epsilon$\\
\hline
{\Gaia} & $G$ & 53 & 1.4 & 0.0 \\
Bialkow & $I$ & 72 & 1.15 & 0.005 \\
APT2 & $I$ & 156 & 1.70 & 0.01 \\
LT & $i$ & 94 & 1.15 & 0.005 \\
DEMONEXT & $I$ & 110 & 1.35 & 0.005 \\
Swarthmore & $I$ & 19 & 1.00 & 0.00 \\
UBT60 & $I$ & 18 & 1.00 & 0.005 \\
ASAS-SN & $V$ & 68 & 1.45 & 0.01 \\
\hline
\end{tabular}
\end{table}

\begin{table}
\caption{Best-fit microlensing model parameters of the Gaia16aye binary event.}
\label{tab:params}
\centering
\begin{tabular}{lr}
\hline \hline
Parameter & Value\\
\hline
$t_0$ (HJD$'$) & $7674.738 \pm 0.057$ \\
$u_0$ & $0.0400 \pm 0.0014$ \\
$t_{\rm E}$ (d) & $111.09 \pm 0.41$ \\
$\pi_{\rm E,N}$ & $-0.373 \pm 0.002$ \\
$\pi_{\rm E,E}$ & $-0.145 \pm 0.001$ \\
$\log\rho$ & $-2.519 \pm 0.003$ \\
$q$ & $0.639 \pm 0.004$ \\
$s_0$ & $1.007 \pm 0.002$ \\
$\alpha$ (rad) & $5.339 \pm 0.002$ \\
$s_z$ & $0.404 \pm 0.028$ \\
$\gamma_x$ (yr$^{-1}$) & $0.384 \pm 0.009$ \\
$\gamma_y$ (yr$^{-1}$) & $0.591 \pm 0.012$ \\
$\gamma_z$ (yr$^{-1}$) & $-1.121 \pm 0.032$ \\
\hline
$I_{\rm s}$ (mag) & $14.70 \pm 0.02$ \\
$I_{\rm blend}$ (mag) & $16.09 \pm 0.02$ \\
$R_{\rm s}$ (mag) & $15.62 \pm 0.02$ \\
$R_{\rm blend}$ (mag) & $17.05 \pm 0.02$ \\
$V_{\rm s}$ (mag) & $16.61 \pm 0.02$ \\
$V_{\rm blend}$ (mag) & $17.98 \pm 0.02$ \\
\hline
\end{tabular}

\begin{footnotesize}
$\mathrm{HJD}'=\mathrm{HJD}-2450000$. We adopt $t_{\rm 0,par} = t_{\rm 0,kep} = 7675$.
\end{footnotesize}
\end{table}

\begin{table}
\caption{Physical parameters of the binary lens system.}
\label{tab:physical}
\centering
\begin{tabular}{lr}
\hline \hline
Parameter & Value\\
\hline
$\theta_{\rm E}$ (mas) & $3.04 \pm 0.24$\\
$\mu_{\rm rel}$ (mas/yr) & $10.1 \pm 0.8$\\
$M_1$ ($M_{\odot}$) & $0.57 \pm 0.05$ \\
$M_2$ ($M_{\odot}$) & $0.36 \pm 0.03$\\
$D_l$ (pc) & $780 \pm 60$\\
\hline
$a$ (au) & $1.98 \pm 0.03$\\
$P$ (yr) & $2.88 \pm 0.05$\\
$e$ & $0.30 \pm 0.03$\\
$i$ (deg) & $65.5 \pm 0.7$\\
$\Omega$ (deg) & $-169.4 \pm 0.9$\\
$\omega$ (deg) & $-30.5 \pm 3.8$\\
$t_{\rm peri}$ (HJD$'$) & $8170 \pm 14$\\
\hline
\end{tabular}

\begin{footnotesize}
Uncertainties of orbital parameters do not include the uncertainty in $\theta_*$ and $D_s$. We adopt $\theta_* = 9.2$\,$\mu$as and $D_s=15$\,kpc.
\end{footnotesize}
\end{table}

\begin{figure*}
\centering
\includegraphics[width=\textwidth]{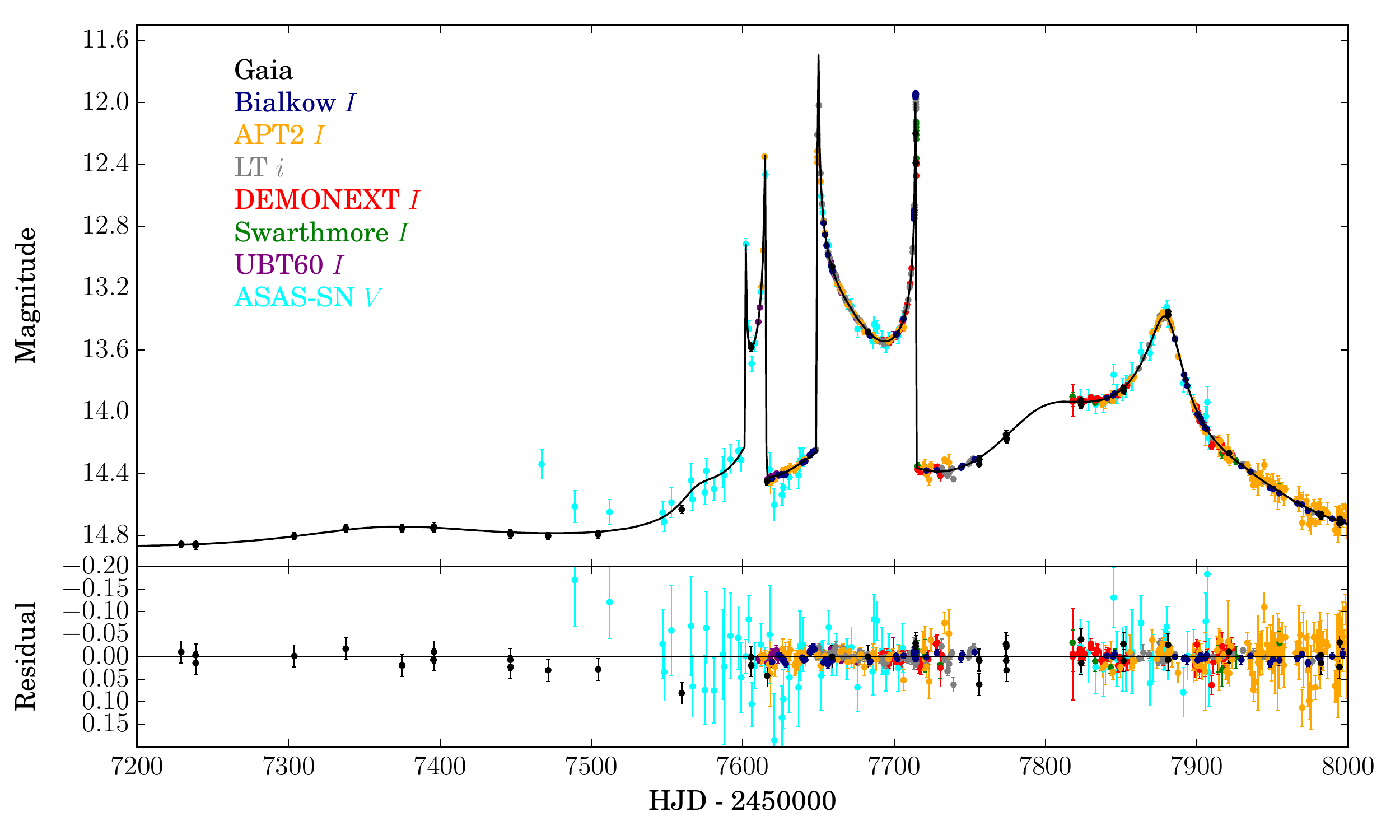}
\caption{Light curve of the microlensing event Gaia16aye, showing only the data used in the microlensing model. All measurements are transformed into the LT $i$-band magnitude scale.}
\label{fig:lc1}
\end{figure*}

\begin{figure*}
\centering
\includegraphics[width=0.9\textwidth]{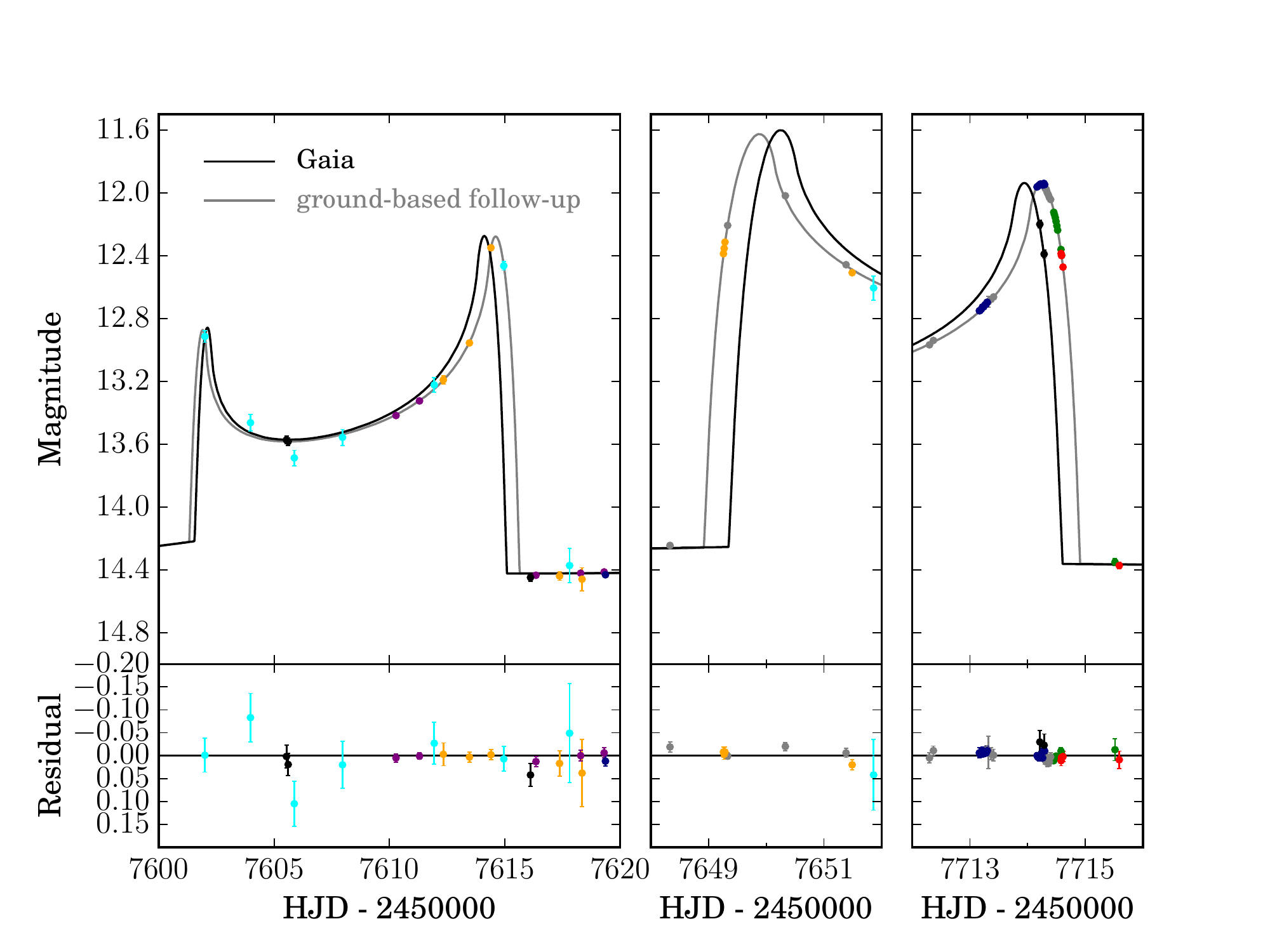}
\caption{Space-based parallax in Gaia16aye. As \textit{Gaia} is separated by 0.01\,au from the Earth, the \textit{Gaia} light curve (black) differs slightly from Earth-based observations (grey curve). Space parallax can be measured through two fortuitous \textit{Gaia} data points collected near $\mathrm{HJD}'\sim7714$. All measurements are transformed into the LT $i$-band magnitude scale. }
\label{fig:lc2}
\end{figure*}

\begin{figure}
\centering
\includegraphics[width=0.95\columnwidth]{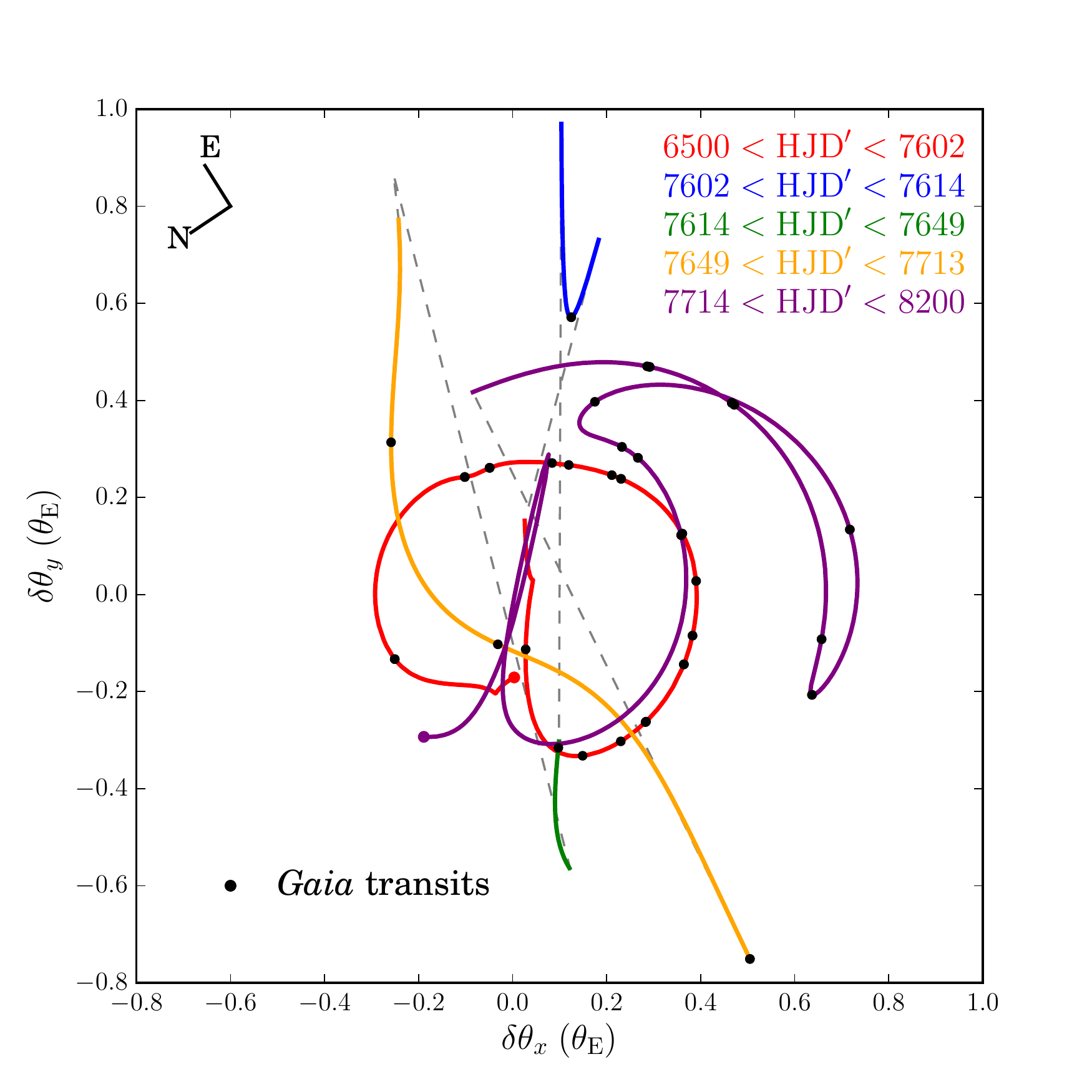}
\caption{As the source star moves across the caustics, new images of the source can be created while others may disappear, resulting in changes of the image centroid. Colour curves show the path of the centroid of the source images relative to the unlensed position of the source (additional light from components of the lens is not included). Moments of \textit{Gaia} transits are marked with black points. The coordinate system is the same as in Fig. \ref{fig:caustic}. The shifts are scaled to the angular Einstein radius of the system ($\theta_{\rm E}=3.04\pm0.24$\,mas). Analysis of the {\Gaia} astrometric measurements will provide an independent estimate of $\theta_{\rm E}$.}
\label{fig:astrometry}
\end{figure}

\section{Discussion}

A massive follow-up campaign allowed us to collect a very detailed light curve for Gaia 16aye and hence to cover the evolution of the event exhaustively. 
Photometric data were obtained over a period of more than two years by a network of observers scattered around the world. It should be emphasised that the vast majority of the observations were taken by enthusiastic individuals, including both professional astronomers and amateurs, who devoted their telescope time to this task. 

The case of Gaia16aye illustrates the power of coordinated long-term time-domain observations, which lead to a scientific discovery. 
The field of microlensing has particularly well benefit in the past from such follow-up observations, which resulted, for example, in the first microlensing planetary discoveries \citep[\eg][]{Udalski2005, Beaulieu2006}. 
This event also offered excitement with its multiple, rapid, and often dramatic changes in brightness. 
Therefore it was also essential to use tools that facilitated the observations and data processing. 
Of particular importance was the Cambridge Photometric Calibration Server (CPCS, \citealt{Zielinski2019}), which performed the standardisation of the photometric observations collected by a large variety of different instruments. Moreover, the operation of the CPCS can be scripted, hence the observations could be automatically uploaded and processed without any human intervention. This solution helped track the evolution of the light curve, especially at times when the event changed dramatically. The processed observations and photometric measurements were immediately available for everyone to view, and appropriate actions were undertaken, such as an increase of the observing cadence when the peak at the fourth caustic crossing was approached. 
We note that no archival catalogues are available in $I$ and $R$ filters for the part of the sky with the Gaia16aye event.
All the observations carried out in these filters were automatically adjusted by the CPCS to the nearest Sloan $i$ and $r$ bands. 
This does not affect the microlensing modelling, but the standardised light curve in $i$ and $r$ filters is systematically offset. On the other hand, the $B-$, $g-$ and $V-$band observations processed by the CPCS are calibrated correctly to the 1\%\ level. 

In the case of Gaia16aye, the light curve contains multiple features, which allowed us to constrain the microlensing model uniquely, despite its complexity. 
In addition to the four caustic crossings and a cusp approach, the microlensing model also predicted a smooth low-amplitude long-term bump about a year before the first caustic crossing, at about HJD'=7350. This feature was indeed found in the {\Gaia} data, see Fig.\ref{fig:lc1}.
The amplitude of this rise was about 0.1 mag, which is close to the level of {\Gaia}'s photometric error bars, and the signal was far too faint to trigger an alert.

Additional confirmation of the correctness of the microlensing model comes from the detection of the microlensing space-parallax effect, see Fig.\ref{fig:lc2}. The offset in the timing of the fourth caustic crossing as seen by {\Gaia} and ground-based telescopes is due to the distance of {\Gaia} of 1.5 million km away from Earth. The offset in time was 6.63h (\ie the caustic crossing by the source occurred first at {\Gaia}'s location) and the amplification difference was  -0.007 mag, that is, it was brighter at {\Gaia}. The model from ground-based data only predicted these offsets to within 3 minutes and 0.003 mag, respectively. This indicates our model is unique and robust.

From the microlensing light curve analysis, we can derive an upper limit on the amount of light emitted by the lensing object, or constraints on the dark nature of the lens can be obtained (\eg \citealt{Yee2015flux,Wyrzykowski2016}).
We find that the masses of the lens components are $0.57\pm0.05\,M_{\odot}$ and $0.36\pm0.03\,M_{\odot}$ and that the lens is located about $D_l = 780\pm60$\,pc from the Sun. Because the $V$-band absolute magnitudes of main-sequence stars of these masses are 8.62 and 11.14 \citep{pecaut2013}, respectively, the total brightness of the binary is $V=17.97$ and $I=16.26$, assuming conservatively $A_V=0.1$ towards the lens. This is consistent with the brightness and colour of the blend ($V_{\rm blend}=17.98$ and $I_{\rm blend}=16.09$). The blended light therefore comes from the lens, which is also consistent with the lack of any additional sources of light on the Keck AO image. This is an additional check that our model is correct.

The largest uncertainty in our lens mass determination comes from the $\theta_\mathrm{E}$ parameter, which we derived from the finite source effects.
Through the multiple caustic crossings, but particularly through very detailed coverage of the fourth crossing with multiple observatories, we were able to constrain the size of the source stellar disc in units of the Einstein radius ($\log \rho$) with an uncertainty smaller than 1\%. However, in order to derive $\theta_\mathrm{E}$, we relied on the colour-angular size relation and theoretical predictions for the de-reddened colour of the source based on its spectral type.
These may have introduced systematic errors to the angular size and hence to the lens mass measurement. 
We also note that the amount of the extinction derived based on our photometry ($A_V=0.62$ mag) is significantly smaller than that measured by \cite{Schlafly2011} in this direction ($A_V=1.6$ mag). This and the uncertainty in the physical size of giant stars affects the estimate of the source distance, but because the lens is very nearby at less than 1 kpc, the source distance does not affect the overall result of this study.

Nevertheless, an independent measurement of the Einstein radius, and thus the final confirmation of the nature of the lens in Gaia16aye, can be obtained in the near future from {\Gaia} astrometric time-domain data.
%
Using our photometry-based model, we computed the positions and amplifications of the images throughout the evolution of the event. 
Figure \ref{fig:astrometry} shows the expected position of the combined light of all the images shown in the frame of the centre of mass of the binary and in units of the Einstein radius. 
The figure shows only the centroid motion due to microlensing relative to the unlensed position of the source.
The moments of {\Gaia} observations are marked with black dots. 
Because $\theta_E=$3.04$\pm$0.24 mas, the expected amplitude of the astrometric variation is about 3 mas.
This should be detectable in {\Gaia} astrometric time-series because {\Gaia} is expected to have error bars in the along-scan direction of about 0.1 mas \citep{Rybicki2018}.
The estimate of $\theta_{\rm E}$ from {\Gaia} will be free of our assumptions about the intrinsic colours of the source and the interstellar extinction.
The actual {\Gaia} astrometry will also include the effects of parallax and proper motion of the source as well as the blended light from both components of the binary lens. 
The contribution of the lens brightness to the total light is about 25\%, therefore the astrometric data might also be affected by the orbital motion of the binary. 
It is worth emphasising that without the microlensing model presented above, obtained from photometric Gaia and follow-up data alone, the interpretation of the Gaia astrometry will not be possible due to the high complexity of the centroid motion.

Radial velocity measurements of nearby binary lenses offer an additional way for post-event verification of the orbital parameters inferred from the microlensing model. 
So far, such an attempt was successfully achieved only in the case of OGLE-2009-BLG-020, a binary lens event with a clear orbital motion effect \citep{Skowron2011}. 
Follow-up observations from the Keck and Magellan telescopes measured the radial velocity of the binary to agree with the one predicted based on the microlensing event full binary lens orbit solution \citep{Yee2016RV}.
The binary system presented in this work (to be denoted Gaia16aye-L, with its components Gaia16aye-La and Gaia16aye-Lb) is nearby ($780\pm60$\,pc) and fairly bright (I$\sim$16.5 mag without the source star), hence such observations are obtainable. The expected amplitude of the radial velocity curve of the primary is about K$\approx$ 7.6 km/s. We strongly encourage such observations to be carried out in order to verify the binary solution found in microlensing. 

Yet another possibility to verify the model might come from AO or other high-resolution imaging techniques \citep[\eg][]{Scott2019} in some years when the source and the lens separate \citep[\eg][]{Jung2018}. With the relative proper motion of 10.1$\pm$0.8  $\mathrm{mas\,yr}^{-1}$, the binary lens should become visible at a separation of about 50 mas even in 2021.

\section{Conclusions}
We analysed the long-lasting event Gaia16aye, which exhibited four caustic crossings and a cusp approach, as well as space-parallax between the Earth and the {\it Gaia} spacecraft.
The very well-sampled light curve allowed us to determine the masses of the binary system (0.57$\pm$0.05 \MSun \ and 0.36$\pm$0.03 \MSun) and all its orbital components. 
We derived the period (2.88$\pm$0.05 years) and semi-major axis (1.98$\pm$0.03 au), as well as the eccentricity of the orbit (0.30$\pm$0.03).
Gaia16aye is one of only a few microlensing binary systems with a full orbital solution, which offers an opportunity for confirming the binary parameters with radial velocity measurements and high-resolution imaging after some years.  
This event will also be detectable as an astrometric microlensing event in the forthcoming {\Gaia} astrometric time-series data. 

Increasingly more such events will be detectable in the current era of large-scale photometric surveys (\eg {\it Gaia}, OGLE, ZTF).
With the forthcoming thousands of alerts from all over the sky with the Large Synoptic Survey Telescope (LSST), it will become a necessity to use automated tools for transients discovery, their follow-up and follow-up data processing in order to fully identify and characterise the most interesting events. 
Robotic observations of selected alerts, automated analysis of the follow-up data, and light curve generation will soon become new standards in transient time-domain astronomy.
The case of Gaia16aye shows that microlensing can be a useful tool for studying also binary systems where the lensing is caused by dark objects. A detection of a microlensing binary system composed of black holes and neutron stars would provide  information about this elusive population of remnants that is complementary to other studies.

\begin{acknowledgements}


This work relies on the results from the European Space Agency (ESA) space mission \Gaia. 
{\Gaia} data are being processed by the {\Gaia} Data Processing and Analysis Consortium (DPAC). 
Funding for the DPAC is provided by national institutions, in particular the institutions participating in the {\Gaia} Multi-Lateral Agreement (MLA). 
The {\Gaia} mission website is \href{https://www.cosmos.esa.int/gaia}{https://www.cosmos.esa.int/gaia}. 
In particular we acknowledge {\Gaia} Photometric Science Alerts Team, website \href{http://gsaweb.ast.cam.ac.uk/alerts}{http://gsaweb.ast.cam.ac.uk/alerts}.

We thank the members of the OGLE team for discussions and support. 
We also would like to thank the Polish Children Fund (KFnRD) for support of an internship of their pupils in Ostrowik Observatory of the Warsaw University, during which some of the data were collected, in particular we thank: Robert Nowicki, Micha{\l} Por\k{e}bski and Karol Niczyj.

The work presented here has been supported by the following grants from the Polish National Science Centre (NCN):
HARMONIA NCN grant 2015/18/M/ST9/00544, 
OPUS NCN grant 2015/17/B/ST9/03167, 
DAINA NCN grant 2017/27/L/ST9/03221, as well as 
European Commission's FP7 and H2020 OPTICON grants (312430 and 730890), 
Polish Ministry of Higher Education support for OPTICON FP7, 3040/7.PR/2014/2,
MNiSW grant DIR/WK/2018/12.
PMr and JS acknowledge support from MAESTRO NCN grant 2014/14/A/ST9/00121 to Andrzej Udalski.

We would like to thank the following members of the AAVSO for their amazing work with collecting vast amounts of data: 
Teofilo Arranz,
James Boardman,
Stephen Brincat,
Geoff Chaplin,
Emery Erdelyi,
Rafael Farfan,
William Goff,
Franklin Guenther,
Kevin Hills,
Jens Jacobsen,
Raymond Kneip,
David Lane,
Fernando Limon Martinez,
Gianpiero Locatelli,
Andrea Mantero,
Attila Madai,
Peter Meadows,
Otmar Nickel,
Arto Oksanen,
Luis Perez,
Roger Pieri,
Ulisse Quadri,
Diego Rodriguez Perez,
Frank Schorr,
George Sjoberg,
Andras Timar,
Ray Tomlin,
Tonny Vanmunster,
Klaus Wenzel,
Thomas Wikander.

We also thank the amateur observers from around the world, in particular, Pietro Capuozzo, Leone Trascianelli, Igor Zharkov from Ardingly College and Angelo Tomassini, Karl-Ludwig Bath. We also thank Roger Pickard from the British Astronomical Association and Matthias Penselin from the German Haus der Astronomie association for their contributions. 

KS thanks Dr. Dmitry Chulkov and Dr. Panagiotis Gavras for the 
interesting discussion of stellar multiplicity.

We acknowledge support of DDT programmes SW2016b12 (WHT) and A34DDT3 (TNG).
The INT, TNG and WHT are operated on the island of La Palma by the Isaac Newton Group of Telescopes in the Spanish Observatorio del Roque de los Muchachos of the Instituto de Astrofisica de Canarias.

The Liverpool Telescope is operated on the island of La Palma by Liverpool John Moores University in the Spanish Observatorio del Roque de los Muchachos of the Instituto de Astrofisica de Canarias with financial support from the UK Science and Technology Facilities Council.
SJF would like to thank the UCL students who assisted with the collection and checking of UCLO data for the observing campaign:
Martina Aghopian, Ashleigh Arendt, Artem Barinov, Luke Barrett, Jasper Berry-Gair, Arjun Bhogal, Charles Bowesman, William Boyd, Andrei Cuceu, Michael Davies, Max Freedman, Gabriel Fu, Abirami Govindaraju, Iandeep Hothi, Clara Matthews Torres, Darius Modirrousta-Galian, Petru Neague, George Pattinson,
Xiaoxi Song, and Brian Yu.
P.Mr. acknowledges support from the Foundation for Polish Science (Program START) and the National Science Center, Poland (grant ETIUDA 2018/28/T/ST9/00096).
AC, AG and NI acknowledge the financial support from the Slovenian Research Agency (research core funding No. P1-0031 and project grant No. J1-8136) and networking support by the COST Action GWverse CA16104.
Skinakas Observatory is a collaborative project of the University of Crete and the Foundation for Research and Technology-Hellas.
Work  by  C.H. was supported by the grant (2017R1A4A1015178) of National Research Foundation of Korea. 
KW acknowledges funding from STFC, and thanks the University of Leicester for the investment in instrumentation. 
We gratefully acknowledge financial support by the European Space Agency under the NELIOTA program, contract No. 4000112943. This work has made use of data obtained with the Kryoneri Prime Focus Instrument, developed by the European Space Agency NELIOTA project on the 1.2 m Kryoneri telescope, which is operated by IAASARS, National Observatory of Athens, Greece.
The Aristarchos telescope is operated on Helmos Observatory by the IAASARS of the National Observatory of Athens.
This work was supported by the GROWTH project funded by the National Science Foundation under Grant No 1545949.
This work was supported by the MINECO (Spanish Ministry of Economy) through grant ESP2016-80079-C2-1-R (MINECO/FEDER, UE) and ESP2014-55996-C2-1-R (MINECO/FEDER, UE) and MDM-2014-0369 of ICCUB (Unidad de Excelencia a Mar\'ia de Maeztu). 
  
This work was supported by the MINECO (Spanish Ministry of Economy) through grant ESP2016-80079-C2-1-R and RTI2018-095076-B-C21 (MINECO/FEDER, UE), and MDM-2014-0369 of ICCUB (Unidad de Excelencia 'Mar\'ia de Maeztu'). The Joan Oro Telescope (TJO) of the Montsec Astronomical Observatory (OAdM) is owned by the Catalan Government and is operated by the Institute for Space Studies of Catalonia (IEEC).

Support to this study has been provided by Agenzia Spaziale Italiana (ASI) through grants ASI I/058/10/0 and ASI 2014-025-R.1.2015.

KW thanks Dipali Thanki and Ray McErlean for their technical support of the scientific programme of the University of Leicester observatory. This work was supported by Royal Society Research Grant RG170230.

CCN thanks the funding from Ministry of Science and Technology (Taiwan) under the contracts 104-2112-M-008-012-MY3 and 104-2923-M-008-004-MY5.

The research leading to these results has received funding from the European Research Council under the European Union's Seventh Framework Programme (FP/2007-2013) / ERC Grant Agreement n.\,320964 (WDTracer).

We thank the Las Cumbres Observatory and its staff for its continuing support of the  project. ASAS-SN is supported by the Gordon and Betty Moore Foundation through grant GBMF5490 to the Ohio State University and NSF grant AST-1515927. Development of ASAS-SN has been supported by NSF grant AST-0908816, the Mt. Cuba Astronomical Foundation, the Center for Cosmology
and AstroParticle Physics at the Ohio State University, the Chinese Academy of Sciences South America Center for Astronomy (CAS- SACA), the Villum Foundation, and George Skestos.
ARM acknowledges support from the MINECO under the Ram\'on y Cajal programme (RYC-2016-20254) and the AYA2017-86274-P grant, and the AGAUR grant SGR-661/2017.
We acknowledge support from the Science and Technology Facilities Council (TB and RWW; ST/P000541/1).
K.Horne acknowledges support from STFC consolidated grant ST/M001296/1.
This work was partly supported by the Research Council of Lithuania, grant No. S-LL-19-2
Authors thank to T\"UB\.ITAK, IKI, KFU, and AST for partial supports in using RTT150 (Russian-Turkish 1.5-m telescope in Antalya).
This work was partially funded by the subsidy 3.6714.2017/8.9 allocated to Kazan Federal University for the state assignment in the sphere of scientific activities.
This research was partially supported by contract DN 18/13-12.12.2017
with the National Science Fund (Bulgaria).
Work by YS was supported by an appointment to the NASA Postdoctoral Program at the Jet Propulsion Laboratory,
California Institute of Technology, administered by Universities Space Research Association
through a contract with NASA.
GD gratefully acknowledges the observing grant support from the Institute
of Astronomy and NAO Rozhen, BAS, via bilateral joint research project ''Study of ICRF radio-sources
and fast variable astronomical objects'' (PI:G.Damljanovic). This work is a part of the
Projects no. 176011 ''Dynamics and kinematics of celestial bodies and systems'', no. 176004 ''Stellar
physics'' and no. 176021 ''Visible and invisible matter in nearby galaxies: theory and observations''
supported by the Ministry of Education, Science and Technological Development of the Republic of
Serbia.
YT acknowledges the support of DFG priority program SPP 1992 ''Exploring the diversity of Extrasolar Planets'' (WA 1074/11-1).
This work of PMi, DM and ZK was supported by the NCN grant no. 2016/21/B/ST9/01126.
ARM acknowledges support from the MINECO Ram{\'o}n y Cajal programme RYJ-2016-20254 and grant AYA\-2017-86274-P and from the AGAUR grant SGR-661/2017.
The work by C.R. was supported by an appointment to the NASA Postdoctoral Program at the Goddard Space Flight Center, administered by USRA through a contract with NASA.

The Faulkes Telescope Project is an education partner of Las Cumbres Observatory (LCO). The Faulkes Telescopes are maintained and operated by LCO.

This research was made possible through the use of the AAVSO Photometric All-Sky Survey (APASS), funded by the Robert Martin Ayers Sciences Fund and NSF AST-1412587.
The Pan-STARRS1 Surveys (PS1) and the PS1 public science archive have been made possible through contributions by the Institute for Astronomy, the University of Hawaii, the Pan-STARRS Project Office, the Max-Planck Society and its participating institutes, the Max Planck Institute for Astronomy, Heidelberg and the Max Planck Institute for Extraterrestrial Physics, Garching, The Johns Hopkins University, Durham University, the University of Edinburgh, the Queen's University Belfast, the Harvard-Smithsonian Center for Astrophysics, the Las Cumbres Observatory Global Telescope Network Incorporated, the National Central University of Taiwan, the Space Telescope Science Institute, the National Aeronautics and Space Administration under Grant No. NNX08AR22G issued through the Planetary Science Division of the NASA Science Mission Directorate, the National Science Foundation Grant No. AST-1238877, the University of Maryland, Eotvos Lorand University (ELTE), the Los Alamos National Laboratory, and the Gordon and Betty Moore Foundation.
Some of the data presented herein were obtained at the W. M. Keck Observatory, which is operated as a scientific partnership among the California Institute of Technology, the University of California and the National Aeronautics and Space Administration. The Observatory was made possible by the generous financial support of the W. M. Keck Foundation.

%

\end{acknowledgements}

\bibliography{refs_Gaia16aye}
\bibliographystyle{aa}

\begin{appendix}
\section{Parameters of the telescopes taking part in the follow-up}
Table \ref{tab:telescopes-instruments} lists the instruments that were used in all telescopes that took part in the photometric follow-up of the Gaia16aye binary microlensing event. 

\begin{table*}
\caption{Photometric instruments used in the follow-up observations of Gaia16aye.}             
\label{tab:telescopes-instruments}      
\centering                          
\begin{tabular}{l c l c}
\hline\hline                 
Telescope code  &  Mirror size [m] & Instrument & Pixel scale [arcsec]  \\ 
\hline                        
AAVSO & -- & -- & -- \\
Akeno50 & 0.5 & 3 x Apogee Alta U6 & 1.64 \\
APT2 & 0.8 & e2v CCD230-42 & 0.93 \\
Aries130 & 1.30 & CCD Andor DZ436 & 0.54 \\
Aristarchos & 2.3 & VersArray 2048B & 0.16 \\
ASASSN & 0.14 & FLI ProLine230 & 7.80 \\
ASV1 & 0.6 & SBIG ST10 XME & 0.23 \\
 & & Apogee Alta E47 & 0.45 \\
ASV2 & 1.4 & Apogee Alta U42 & 0.24 \\
AUT25 & 0.25 & QSI532swg & 0.71 \\
BAS2 & 2.0 & CCD VersArray 1300B & 0.74 \\
 & & Photometrics for FoReRo2 system & 0.88 \\
BAS50/70 & 0.5/0.7 & FLI ProLine16803 & 1.08 \\
Bialkow & 0.6 & Andor iKon DW432-BV & 0.61 \\
C2PU & 1.04 & SBIG ST16803 & 0.56 \\
Conti & 0.28 & SX694 mono CCD & 0.56 \\
CrAO & 0.2 & SBIG ST8300M & 1.10 \\
DEMONEXT & 0.5 & Fairchild CCD3041 2k x 2k array & 0.90 \\
Foligno & 0.3 & Nikon D90 & 0.76 \\
HAO50 & 0.5 & ATIK314+ & 0.67 \\
Krakow50 & 0.5 & Apogee Alta U42 & 0.42 \\
Kryoneri & 1.2 & Andor Zyla 5.5 & 0.40 \\
LCO-Texas & 1.0 & Sinistro 4k x 4k & 0.39 \\
LCO-Hawaii & 0.4 & SBIG STL-6303 3k x 2k & 1.14 \\
 & 2.0 & Spectral 4k x 4k & 0.30 \\
Leicester & 0.5 & SBIG ST2000XM (before 2017 Nov) & 0.89 \\ 
 & & Moravian G3-11000 (after 2017 Nov) & 1.08 \\
Loiano & 1.52 & BFOSC & 0.58 \\
LOT1m & 1.0 & Apogee Alta U42 & 0.35 \\
LT & 2.0 & IO:O e2v CCD231 & 0.27 \\
MAO165 & 1.65 & Apogee Alta U47 & 0.51 \\
Mercator & 1.2 & Merope & 0.19 \\
Montarrenti & 0.53 & Apogee Alta U47 & 1.16 \\ 
OHP & 1.2 & 1k x 1k CCD & 0.67 \\
OndrejovD50 & 0.5 & CCD FLI IMG 4710 & 1.18 \\
Ostrowik & 0.6 & CCD 512 x 512 Tektronix & 0.76 \\
PIRATE & 0.42 & FLI ProLine16803 & 0.63 \\
pt5m & 0.5 & QSI532 CCD & 0.28 \\
RTT150 & 1.5 & TFOSC & 0.39 \\
SAI & 0.6 & Apogee Aspen CG42 & 0.76 \\
Salerno & 0.6 & FLI ProLine230 & 0.60 \\
SKAS-KFU28 & 0.28 & QSI 583wsg & 0.40 \\
Skinakas & 1.3 & Andor DZ436 & 0.28 \\
SKYNET & 1.0 & 512 x 512 CCD 48um & 1.21 \\
Swarthmore24 & 0.6 & Apogee Alta U16M & 0.38 \\
T60 & 0.6 & FLI ProLine3041 & 0.51 \\
T100 & 1.0 & 4k x 4k CCD & 0.31 \\
TJO & 0.8 & MEIA e2V CCD42-40 & 0.36 \\
TRT-GAO & 0.7 & Andor iKon-L 936 & 0.61 \\
TRT-TNO & 0.5 & Andor iKon-L 936 & 0.68 \\
UCLO-C14E & 0.35 & SBIG STL6303E & 0.86 \\
UCLO-C14W & 0.35 & SBIG STL6303E & 0.86 \\
UBT60 & 0.6 & Apogee Alta U47 & 0.68 \\
Watcher & 0.4 & Andor iXon EM+ & 0.60 \\
WHT-ACAM & 4.2 & ACAM & 0.25 \\
Wise1m & 1.0 & PI camera & 0.58 \\
WiseC28 & 0.71 & FLI ProLine16801 & 0.83 \\

\hline                                   
\end{tabular}
\end{table*}

\section{}
Table \ref{tab:gaiaphot} contains all {\Gaia} mean G-band photometry for the Gaia16aye event that was collected and calibrated by the {\Gaia} Science Alerts system, available at the webpage  \href{http://gsaweb.ast.cam.ac.uk/alerts/alert/Gaia16aye/}{http://gsaweb.ast.cam.ac.uk/alerts/alert/Gaia16aye}. The typical error bar is about 0.1 mag.

\begin{table}
\caption{Gaia photometric measurements of the Gaia16aye microlensing event. The full table is  available in the electronic form of the article. TCB is the barycentric coordinate time.}             
\centering                          
\begin{scriptsize}
\label{tab:gaiaphot}
\begin{tabular}{c c c}        
\hline\hline                 
\multicolumn{2}{c}{Observation date} & average  \\
TCB & JD        & G mag \\
\hline                        
2014-10-30 20:50:59 & 2456961.369 & 15.48 \\
2014-10-30 22:37:33 & 2456961.443 & 15.48 \\
2015-02-15 09:54:03 & 2457068.913 & 15.44 \\
2015-02-15 14:07:43 & 2457069.089 & 15.44 \\
2015-02-15 15:54:18 & 2457069.163 & 15.45 \\
2015-03-09 08:16:20 & 2457090.845 & 15.45 \\
2015-03-09 10:02:55 & 2457090.919 & 15.43 \\
2015-03-09 14:16:35 & 2457091.095 & 15.45 \\
2015-03-09 16:03:10 & 2457091.169 & 15.45 \\
2015-05-20 19:20:37 & 2457163.306 & 15.45 \\
2015-06-10 03:08:39 & 2457183.631 & 15.47 \\
2015-07-25 13:45:22 & 2457229.073 & 15.45 \\
2015-08-04 00:05:24 & 2457238.504 & 15.45 \\
2015-08-04 01:51:58 & 2457238.578 & 15.46 \\
2015-10-08 06:23:08 & 2457303.766 & 15.40 \\
2015-11-11 05:44:30 & 2457337.739 & 15.35 \\
2015-12-18 09:29:34 & 2457374.896 & 15.35 \\
2015-12-18 11:16:08 & 2457374.970 & 15.35 \\
2016-01-08 03:37:06 & 2457395.651 & 15.35 \\
2016-01-08 05:23:40 & 2457395.725 & 15.35 \\
2016-01-08 09:37:20 & 2457395.901 & 15.39 \\
2016-01-08 11:23:54 & 2457395.975 & 15.34 \\
2016-02-27 21:18:55 & 2457446.388 & 15.48 \\
2016-02-27 23:05:29 & 2457446.462 & 15.38 \\
2016-02-28 03:19:09 & 2457446.638 & 15.39 \\
2016-03-23 23:08:54 & 2457471.465 & 15.40 \\
2016-04-25 22:50:35 & 2457504.452 & 15.39 \\
2016-06-02 20:18:57 & 2457542.346 & 15.52 \\
2016-06-20 04:10:13 & 2457559.674 & 15.23 \\
2016-08-05 00:53:51 & 2457605.537 & 14.18 \\
2016-08-05 02:40:25 & 2457605.611 & 14.19 \\
2016-08-05 06:54:05 & 2457605.788 & 14.40 \\
2016-08-05 08:40:39 & 2457605.862 & 14.25 \\
2016-08-15 13:00:28 & 2457616.042 & 15.26 \\
2016-08-15 14:47:02 & 2457616.116 & 15.05 \\
2016-09-27 13:28:36 & 2457659.062 & 13.67 \\
2016-10-21 05:33:20 & 2457682.731 & 14.09 \\
2016-11-21 17:05:46 & 2457714.212 & 12.81 \\
2016-11-21 18:52:20 & 2457714.286 & 13.00 \\
2017-01-02 12:24:22 & 2457756.017 & 14.91 \\
2017-01-02 16:38:01 & 2457756.193 & 14.94 \\
2017-01-02 18:24:35 & 2457756.267 & 14.91 \\
2017-01-20 10:48:21 & 2457773.950 & 14.75 \\
2017-01-20 12:34:55 & 2457774.024 & 14.77 \\
2017-01-20 16:48:35 & 2457774.200 & 14.75 \\
2017-01-20 18:35:09 & 2457774.274 & 14.78 \\
2017-03-10 23:52:28 & 2457823.495 & 14.53 \\
2017-03-11 01:39:02 & 2457823.569 & 14.56 \\
2017-04-07 23:48:22 & 2457851.492 & 14.45 \\
2017-04-08 01:34:57 & 2457851.566 & 14.47 \\
2017-05-07 11:34:44 & 2457880.982 & 13.96 \\
2017-05-07 13:21:19 & 2457881.056 & 13.98 \\
2017-06-16 16:39:01 & 2457921.194 & 14.87 \\
2017-08-16 09:12:15 & 2457981.884 & 15.26 \\
2017-08-16 10:58:49 & 2457981.958 & 15.27 \\
2017-08-28 17:04:45 & 2457994.212 & 15.32 \\
2017-08-28 21:18:24 & 2457994.388 & 15.29 \\
2017-10-08 14:08:21 & 2458035.089 & 15.4 \\
2017-10-08 15:54:55 & 2458035.163 & 15.41 \\
2017-11-04 03:39:50 & 2458061.653 & 15.55 \\
2017-12-03 09:23:18 & 2458090.891 & 15.53 \\
2018-01-18 19:12:05 & 2458137.300 & 15.53 \\
2018-01-18 20:58:40 & 2458137.374 & 15.53 \\
2018-01-19 01:12:20 & 2458137.550 & 15.52 \\
2018-01-19 07:12:33 & 2458137.800 & 15.54 \\
2018-02-04 19:23:34 & 2458154.308 & 15.52 \\
2018-02-04 21:10:08 & 2458154.382 & 15.51 \\
2018-02-05 01:23:49 & 2458154.558 & 15.51 \\
2018-02-05 03:10:23 & 2458154.632 & 15.51 \\
2018-03-23 01:03:21 & 2458200.544 & 15.54 \\
2018-04-22 12:49:53 & 2458231.035 & 15.54 \\
2018-04-22 14:36:27 & 2458231.109 & 15.56 \\
2018-05-19 00:41:48 & 2458257.529 & 15.53 \\
2018-06-30 07:22:25 & 2458299.807 & 15.56 \\
2018-07-12 01:29:24 & 2458311.562 & 15.58 \\
... & ... & \\
\hline                                   
\end{tabular}
\end{scriptsize}
\end{table}

\section{Photometric follow-up data}
Photometric follow-up observations calibrated with the Cambridge Photometric Calibration Server are gathered in table \ref{tab:followupphot}. 
The complete table is available in the electronic form of the article.

\begin{table}
\caption{Photometric follow-up observations of Gaia16aye. ID denotes the unique id of the observation in the Calibration Server.}
\label{tab:followupphot}      
\centering 
\begin{scriptsize}
\begin{tabular}{l l l l l l}
\hline\hline                 
ID & MJD         &  Magnitude & Error & Filter & Observatory/Observer  \\ 
   &  [d] &     [mag]  & [mag] &        & \\
\hline                        
41329 & 57609.74664 & 16.635 & 0.052 & B & UBT60 V.Bakis \\
41348 & 57609.74742 & 14.914 & 0.012 & V & UBT60 V.Bakis \\
41367 & 57609.74819 & 14.108 & 0.006 & r & UBT60 V.Bakis \\
41386 & 57609.74897 & 13.375 & 0.005 & i & UBT60 V.Bakis \\
41330 & 57609.74978 & 16.548 & 0.037 & B & UBT60 V.Bakis \\
41349 & 57609.75055 & 14.907 & 0.010 & V & UBT60 V.Bakis \\
41368 & 57609.75133 & 14.102 & 0.005 & r & UBT60 V.Bakis \\
41387 & 57609.75210 & 13.378 & 0.005 & i & UBT60 V.Bakis \\
41331 & 57609.75281 & 16.600 & 0.037 & B & UBT60 V.Bakis \\
41350 & 57609.75359 & 14.897 & 0.010 & V & UBT60 V.Bakis \\
41369 & 57609.75436 & 14.117 & 0.005 & r & UBT60 V.Bakis \\
41388 & 57609.75514 & 13.374 & 0.005 & i & UBT60 V.Bakis \\
41332 & 57609.75588 & 16.504 & 0.035 & B & UBT60 V.Bakis \\
41351 & 57609.75665 & 14.902 & 0.010 & V & UBT60 V.Bakis \\
41370 & 57609.75743 & 14.105 & 0.005 & r & UBT60 V.Bakis \\
41389 & 57609.75820 & 13.399 & 0.005 & i & UBT60 V.Bakis \\
41333 & 57609.75896 & 16.538 & 0.035 & B & UBT60 V.Bakis \\
41352 & 57609.75973 & 14.904 & 0.010 & V & UBT60 V.Bakis \\
41371 & 57609.76051 & 14.117 & 0.006 & r & UBT60 V.Bakis \\
41390 & 57609.76128 & 13.403 & 0.005 & i & UBT60 V.Bakis \\
54690 & 57609.78240 & 14.202 & 0.009 & r & SAI A.Zubareva \\
54689 & 57609.78569 & 16.528 & 0.024 & B & SAI A.Zubareva \\
54680 & 57609.78902 & 16.544 & 0.016 & B & SAI A.Zubareva \\
54663 & 57609.79078 & 14.974 & 0.007 & V & SAI A.Zubareva \\
54681 & 57609.79218 & 14.148 & 0.005 & r & SAI A.Zubareva \\
54682 & 57609.79395 & 16.539 & 0.019 & B & SAI A.Zubareva \\
41334 & 57609.79522 & 16.599 & 0.024 & B & UBT60 V.Bakis \\
54664 & 57609.79569 & 14.971 & 0.008 & V & SAI A.Zubareva \\
41353 & 57609.79600 & 14.884 & 0.008 & V & UBT60 V.Bakis \\
41372 & 57609.79677 & 14.082 & 0.005 & r & UBT60 V.Bakis \\
54683 & 57609.79711 & 14.167 & 0.005 & r & SAI A.Zubareva \\
41391 & 57609.79755 & 13.355 & 0.005 & i & UBT60 V.Bakis \\
54684 & 57609.79888 & 16.583 & 0.020 & B & SAI A.Zubareva \\
54665 & 57609.80063 & 15.014 & 0.009 & V & SAI A.Zubareva \\
54685 & 57609.80202 & 14.168 & 0.005 & r & SAI A.Zubareva \\
54686 & 57609.80373 & 14.178 & 0.005 & r & SAI A.Zubareva \\
41335 & 57609.80477 & 16.605 & 0.026 & B & UBT60 V.Bakis \\
41354 & 57609.80554 & 14.876 & 0.008 & V & UBT60 V.Bakis \\
41373 & 57609.80632 & 14.102 & 0.005 & r & UBT60 V.Bakis \\
41392 & 57609.80709 & 13.374 & 0.005 & i & UBT60 V.Bakis \\
41336 & 57609.80787 & 16.549 & 0.025 & B & UBT60 V.Bakis \\
41355 & 57609.80864 & 14.864 & 0.008 & V & UBT60 V.Bakis \\
41374 & 57609.80942 & 14.106 & 0.005 & r & UBT60 V.Bakis \\
41393 & 57609.81019 & 13.380 & 0.005 & i & UBT60 V.Bakis \\
41337 & 57609.81094 & 16.488 & 0.025 & B & UBT60 V.Bakis \\
41356 & 57609.81171 & 14.884 & 0.008 & V & UBT60 V.Bakis \\
41375 & 57609.81249 & 14.102 & 0.005 & r & UBT60 V.Bakis \\
41394 & 57609.81326 & 13.382 & 0.005 & i & UBT60 V.Bakis \\
41338 & 57609.81405 & 16.492 & 0.027 & B & UBT60 V.Bakis \\
41357 & 57609.81483 & 14.879 & 0.008 & V & UBT60 V.Bakis \\
41376 & 57609.81560 & 14.101 & 0.005 & r & UBT60 V.Bakis \\
41395 & 57609.81638 & 13.374 & 0.005 & i & UBT60 V.Bakis \\
40186 & 57609.90821 & 17.158 & 0.134 & B & pt5m L.Hardy \\
40187 & 57609.90927 & 16.939 & 0.116 & B & pt5m L.Hardy \\
40188 & 57609.91009 & 16.548 & 0.098 & B & pt5m L.Hardy \\
40189 & 57609.91092 & 14.917 & 0.022 & V & pt5m L.Hardy \\
40190 & 57609.91181 & 14.964 & 0.021 & V & pt5m L.Hardy \\
40191 & 57609.91263 & 14.958 & 0.022 & V & pt5m L.Hardy \\
40192 & 57609.91346 & 14.132 & 0.009 & r & pt5m L.Hardy \\
40193 & 57609.91457 & 14.188 & 0.011 & r & pt5m L.Hardy \\
40194 & 57609.91540 & 14.106 & 0.010 & r & pt5m L.Hardy \\
40195 & 57609.91640 & 13.439 & 0.010 & i & pt5m L.Hardy \\
40196 & 57609.91751 & 13.448 & 0.009 & i & pt5m L.Hardy \\
40197 & 57609.91834 & 13.453 & 0.010 & i & pt5m L.Hardy \\
40268 & 57610.00399 & 16.522 & 0.014 & B & TJO U.Burgaz \\
40271 & 57610.01489 & 15.002 & 0.006 & V & TJO U.Burgaz \\
40272 & 57610.01842 & 14.956 & 0.020 & V & TJO U.Burgaz \\
40274 & 57610.03669 & 13.107 & 0.055 & i & TJO U.Burgaz \\
40275 & 57610.04022 & 13.293 & 0.011 & i & TJO U.Burgaz \\
40276 & 57610.04375 & 13.388 & 0.004 & i & TJO U.Burgaz \\
54687 & 57610.05719 & 16.491 & 0.057 & B & SAI A.Zubareva \\
54666 & 57610.05894 & 14.977 & 0.018 & V & SAI A.Zubareva \\
54688 & 57610.06035 & 14.192 & 0.009 & r & SAI A.Zubareva \\
41339 & 57610.76348 & 16.499 & 0.029 & B & UBT60 V.Bakis \\
41358 & 57610.76424 & 14.805 & 0.009 & V & UBT60 V.Bakis \\
41377 & 57610.76499 & 14.009 & 0.005 & r & UBT60 V.Bakis \\
41396 & 57610.76576 & 13.285 & 0.005 & i & UBT60 V.Bakis \\
...   & ...         & ...    & ...   & . & ... \\
\hline                                   
\end{tabular}
\end{scriptsize}
\end{table}

\section{Photometric data used in the microlensing modelling}
Photometric observations that were used in the microlensing model are shown in table \ref{tab:followupphot-model}. The complete table is available in the electronic form of the article.

\begin{table}
\caption{Photometric follow-up observations of Gaia16aye used in the model. Observatory codes: 1 {\Gaia} (G), 2 Bialkow (I), 3 APT2 (I), 4 LT (i), 5 DEMONEXT (I), 6 Swarthmore (I), 7 UBT60 (I), and 8 ASAS-SN (V). The full data set is available in the on-line version of the paper.} 
\label{tab:followupphot-model}      
\centering 
\begin{scriptsize}
\begin{tabular}{l l l l}
\hline\hline                 
HJD [d]  &  Magnitude [mag] & Error [mag] & Observatory code  \\ 
\hline                        
2456961.36775 & 15.480 & 0.010 & 1\\
2456961.44175 & 15.480 & 0.010 & 1\\
2457068.91154 & 15.440 & 0.010 & 1\\
... & ... & ... & ... \\
2457619.36442 & 14.350 & 0.009 & 2\\
2457623.42542 & 14.323 & 0.006 & 2\\
2457625.43582 & 14.320 & 0.006 & 2\\
... & ... & ... & ... \\
2457612.33545 & 13.127 & 0.013 & 3\\
2457613.46778 & 12.894 & 0.003 & 3\\
2457614.40174 & 12.293 & 0.003 & 3\\
... & ... & ... & ... \\
2457647.43662 & 14.256 & 0.007 & 4\\
2457648.33147 & 14.245 & 0.009 & 4\\
2457649.33125 & 12.208 & 0.004 & 4\\
... & ... & ... & ... \\
2457690.67443 & 13.433 & 0.007 & 5\\
2457691.65978 & 13.433 & 0.006 & 5\\
2457692.59705 & 13.428 & 0.006 & 5\\
... & ... & ... & ... \\
2457714.45266 & 12.246 & 0.003 & 6\\
2457714.46433 & 12.261 & 0.004 & 6\\
2457714.47873 & 12.280 & 0.005 & 6\\
... & ... & ... & ... \\
2457610.28565 & 13.379 & 0.007 & 7\\
2457611.30428 & 13.286 & 0.005 & 7\\
2457616.35217 & 14.400 & 0.010 & 7\\
... & ... & ... & ... \\
2457467.10912 & 17.020 & 0.170 & 8\\
2457489.03978 & 17.940 & 0.330 & 8\\
2457512.02932 & 18.110 & 0.290 & 8\\
... & ... & ... & ... \\
\hline                                   
\end{tabular}
\end{scriptsize}
\end{table}

\end{appendix}

\end{document}